\documentclass[12pt]{article}

\usepackage{amssymb}
\usepackage{amsfonts}
\usepackage{amsmath}
\usepackage{amsthm}
\usepackage{mathtools}
\usepackage{natbib}
\usepackage{bibunits}
\defaultbibliographystyle{abbrvnat}
\defaultbibliography{AnnaBib_SV,complete}

\usepackage[nohead]{geometry}
\usepackage[singlespacing]{setspace}
\usepackage[bottom]{footmisc}
\usepackage{indentfirst}
\usepackage{endnotes}
\usepackage{graphicx}%
\usepackage{rotating}
\usepackage{verbatim}
\usepackage{setspace}
\usepackage{multirow}
\usepackage{latexsym}
\usepackage{bigints}
\usepackage{algorithm}
\usepackage{algpseudocode}
\usepackage{makecell}

\usepackage{makeidx}
\usepackage{fancyhdr}
\usepackage{type1cm}

\usepackage{booktabs}
\usepackage{tabularx}

\usepackage{hyperref}
\hypersetup{
    colorlinks = true,
    citecolor = {blue}
}

\usepackage{booktabs} 

\usepackage{todonotes}

\newcommand{\by}{\mbox{\bf y}}
\newcommand{\EE}{\mbox{\bf E}}

\newcommand{\wh}{\widehat}
\newcommand{\wtl}{\widetilde}

\newcommand{\var}{\mathrm{var}}
\newcommand{\beq}{\begin{eqnarray*}}
\newcommand{\eeq}{\end{eqnarray*}}

\newcommand{\smcsq}{SMC$^2$}  
\newcommand{\bigO}{\mathcal{O}}
\newcommand{\diag}{\operatorname{diag}}

\theoremstyle{definition}

\newtheorem{remark}{Remark}[section]

\usepackage{eqparbox}

\makeatletter
\def\@biblabel#1{\hspace*{-\labelsep}}
\makeatother
\begin{document}
\begin{bibunit}
\title{\huge Bayesian inference, on-line forecasting and model choice for large VAR models with Cholesky stochastic volatility}
\author{Nicolas Chopin\thanks{
CREST, ENSAE, 5 Avenue Henry Le Chatelier, 91120 Palaiseau (France). e-mail:
\texttt{nicolas.chopin@math.cnrs.fr}} \and
Andras Fulop\thanks{ESSEC Business School (France). e-mail:
\texttt{fulop@essec.edu}} \and
Yuedan Huo\thanks{
CREST, ENSAE, 5 Avenue Henry Le Chatelier, 91120 Palaiseau (France). e-mail:
\texttt{yuedan.huo@ensae.fr}} \and
Anna Simoni\thanks{
CREST, CNRS, \'{E}cole Polytechnique, ENSAE, 5 Avenue Henry Le Chatelier, 91120 Palaiseau (France). Phone: +33(0)170266837. e-mail: \texttt{anna.simoni@polytechnique.edu}}
}

\date{}

\maketitle
\begin{abstract}
  We consider $K$-dimensional Bayesian vector autoregressions (BVARs) with
  Cholesky stochastic volatility (SV), in which the innovation covariance
  matrix is a lower-triangular linear transform of $K$ independent univariate
  SV processes. Such models are widely used in empirical macroeconomics to
  capture time-varying uncertainty and improve forecast accuracy, but the cost
  of posterior simulation is the binding constraint on the size of the system,
  and a major bottleneck for empirical work. We introduce a Markov chain Monte
  Carlo (MCMC) kernel that mixes better than existing samplers at the same
  computational complexity. It rests on a reparametrisation that makes the $K$ volatility trajectories conditionally independent, and on Particle Gibbs to
  update each trajectory. The kernel targets the exact posterior, rather than
  an approximation of it, and in an application with $K = 15$ it raises the
  mean effective sample size per second, relative to the benchmark corrected
  triangular algorithm, by a factor of approximately $14$ for the VAR coefficients and
  $3.4$ for the volatilities. We also introduce a a Sequential Monte
  Carlo squared (\smcsq{}) sampler, which used our MCMC kernel as a building
  block,  and  which delivers at every $t$ the one-step-ahead predictive
  density and the marginal likelihood of the data up to $t$, and hence on-line
  forecasting and model choice. To our knowledge, this is the first algorithm
  that delivers sequential marginal likelihoods for Cholesky-SV BVARs with
  static contemporaneous coefficients and a non-conjugate prior. We illustrate
  both on US monthly macroeconomic data, with $K=15$ for posterior inference
  and $K=6$ for the sequential sampler and model choice.
\end{abstract}

\begin{singlespace}

\end{singlespace}

\section{Introduction}\label{sec:intro}

\subsection{Motivation}
Vector autoregressions (VARs) are a standard tool in empirical macroeconomics for structural analysis and forecasting. Since the seminal work of \citet{Sims1980}, which introduced VARs to economics, a large literature has demonstrated that a good specification of VAR models requires two ingredients. The first is a rich conditional mean, with a sizeable cross-section of $K$ variables and a long lag length $p$, which mitigates the misspecification that arises when a small VAR cannot represent the true data generating process. The second is time variation in the volatilities, typically modelled as stochastic volatility (SV). \citet{Clark2011} shows that adding SV to Bayesian VARs materially improves real-time density forecasts (see also \citet{ClarkRavazzolo2015} and the references therein).

Because even small VARs have a large number of parameters, Bayesian shrinkage has a long tradition as a method to handle large unrestricted VAR and
to address the parameter proliferation issue. \citet{BanburaGiannoneReichlin2010} and successive works including \citet{Koop2013} have shown that large Bayesian VAR provide a competitive alternative to factor models to handle large datasets. A large BVAR-SV with many lags accentuates the issues of parameter proliferation and computational burden even more.

In this paper we consider BVAR-SV models based on the Cholesky decomposition \citep{COGLEY2005, Primiceri2005}, that is, the $(K \times K)$-covariance matrix $\Sigma_t$ of the innovation is of the form $\Sigma_t = A^{-1} \Lambda_t (A^\top)^{-1}$, where $A$ is lower triangular with ones on the diagonal, and $\Lambda_t$ is a diagonal matrix that contains the $K$ time-varying volatility processes. \citet{Chan2023} compares stochastic volatility specifications for large Bayesian VARs and finds that both the Cholesky and the factor specifications fit the data better than simpler models based on a common (univariate) stochastic volatility process.

However, while these SV specifications improve model fit and flexibility, sampling from the posterior distribution is computationally intensive because it involves manipulating large matrices. Informative priors effectively tackle the overfitting induced by the large
dimension, but they do not resolve this computational burden. A naive Gibbs
sampler for this problem has complexity $\bigO(TK^6)$, where $T$ is the sample
size. This makes it practically impossible to consider datasets where $K \gg
10$. \citet{CARRIERO2019137} exploit a triangularisation of the system to derive an equation-by-equation kernel with complexity $\bigO(TK^4)$. The corrected version of their algorithm \citep{CarrieroChanClarkMarcellino2022} serves as our baseline.

\subsection{Contributions}
We make two contributions in this paper. First, we derive an MCMC kernel for this class of models and show that it outperforms the corrected triangular algorithm of \citet{CarrieroChanClarkMarcellino2022} in two respects, at the same computational complexity. The first concerns mixing. In our application to a VAR with $K = 15$ monthly macroeconomic series and $p = 9$ lags, our kernel raises the  mean effective sample size per second by a factor of about 14 for the VAR coefficients and about 3.4 for the volatility states. The second improvement is exactness. Our kernel targets the exact posterior of the model, whereas the standard treatment of the volatilities relies on the mixture approximation of \citet{KimShephardChib1998} and on the ordering of the Gibbs steps established by \citet{DelNegroPrimiceri2015}.

Our MCMC approach relies on a reparametrisation that makes the $K$ components of the transformed data conditionally independent. This reparametrisation offers two advantages. First, it strongly reduces the posterior correlation between the blocks updated by the sampler, so the kernel moves more freely in the parameter space. Second, it makes it easy to implement the CSMC (conditional sequential Monte Carlo) methodology of \citet{PMCMC} to sample the $K$ volatility trajectories. Under the reparametrised model, these trajectories are conditionally independent and may be treated as the state trajectories of $K$ univariate SV models, which may moreover be updated in parallel. CSMC updates the states of a state-space model exactly, in the sense that it leaves their exact conditional distribution invariant and so involves no approximation error, and efficiently, in the sense that it mixes well even with a small number of particles. It is an instance of the more general PMCMC (particle MCMC) methodology, which embeds a particle filter inside an MCMC sampler to generate candidates for the hidden variables, here the volatilities.

The second contribution is an SMC$^2$ sampler \citep{smc2, FulopLi} that computes recursively the posterior distribution of the parameters given the data up to time $t$, for $t = 1, 2, \ldots$. The same run delivers, at every $t$, the one-step-ahead predictive density and the marginal likelihood of the data up to time $t$. It therefore performs out-of-sample forecasting and model choice on-line. For instance, we can compare models with different lag lengths as data accrue, and detect changes in their relative performance. As a by-product, the sampler accommodates the ragged edge of real-time data \citep{Wallis1986}, that is, a final period in which the last series in the model's ordering have not yet been released. The updates of the affected equations are simply truncated, and no missing value is imputed, in contrast with the data-augmentation treatment that is standard in nowcasting applications \citep{SchorfheideSong2015, CarrieroClarkMarcellino2015}. To the best of our knowledge, this is the first algorithm that delivers sequential marginal likelihoods for Cholesky-SV VARs with static contemporaneous coefficients and a non-conjugate prior. 

By on-line we mean that these quantities are produced in a single forward pass
through the sample, each observation being assimilated once, when it arrives.
The conventional alternative is to re-estimate the model on each expanding
window $y_{1:t}$, which multiplies the cost by the number of forecast origins
and does not deliver marginal likelihoods as a by-product. We do not mean that the cost per observation is bounded: each rejuvenation step applies the Particle Gibbs kernel to all data accumulated up to that point, so the cost of a move grows with $t$.

The two contributions rest on the same building block, a particle filter for the volatility paths. Run forward, the filter provides unbiased estimates of the likelihood. SMC$^2$ uses these estimates to reweight its parameter particles and to evaluate the marginal likelihood sequentially. Run conditionally on a retained trajectory, the filter becomes the CSMC kernel. Our MCMC sampler uses this kernel to update the volatilities, and SMC$^2$ uses it to rejuvenate its parameter particles. We also show that combining the genealogy-tracking device of \citet{pathstorage} with backward-sampling CSMC and the waste-free scheme of \citet{wastefreeSMC} keeps the memory cost of SMC$^2$ low without sacrificing mixing (Section~\ref{sub:smc2:memory}).


\subsection{Related Literature}\label{sec:related}

Our work connects three strands of the Bayesian VAR literature: the modelling of time-varying volatility, the design of scalable estimation algorithms for
high-dimensional systems, and the application of sequential Monte Carlo methods to VAR models.\\
\indent Estimation of the Cholesky SV models introduced above has conventionally relied on one of two samplers for the log-volatility block: the single-move algorithm of \citet{JacquierPolsonRossi1994}, as in \citet{COGLEY2005}, and the auxiliary mixture sampler of \citet{KimShephardChib1998}, as in \citet{Primiceri2005}. The latter approximates the $\log\chi_1^2$ measurement density by a mixture of Gaussians and thereby restores a conditionally linear and Gaussian state space, and has become the more common choice on efficiency grounds. \citet{DelNegroPrimiceri2015} establish the ordering of the Gibbs steps for this sampler, and this ordering has since become standard. Our MCMC kernel requires neither the mixture approximation nor the associated ordering constraints.\\
\indent Scaling these models to the large systems favoured in modern macroeconometrics \citep{BanburaGiannoneReichlin2010} is challenging because a naive Gibbs sampler manipulates covariance matrices of the full system dimension. Two broad routes have emerged. The first exploits the triangular structure to estimate the system equation-by-equation: \citet{CARRIERO2019137} and \citet{CarrieroChanClarkMarcellino2022} show that this lowers the order of complexity from $\bigO(TK^6)$ to $\bigO(TK^4)$, making datasets with $K \gg 10$ feasible. A related approach achieves equation-by-equation estimation through a factor SV structure, which decouples the equations conditionally on the latent factors, combined with global-local shrinkage priors \citep{KASTNER2019}. The second route replaces Kalman-filter smoothing of the latent states with precision-based (band and sparse matrix) samplers \citep{ChanJeliazkov2009} and, for the coefficient block under shrinkage priors, with the fast structured Gaussian sampler of \citet{BhattacharyaChakrabortyMallick2016}.\\
\indent Our kernel belongs to the first route. Its reparametrisation, the recursive structural form of the VAR, is not new: it underlies the sampler of \citet{WaggonerZha2003} and several samplers for large BVAR-SV models \citep{Chan2022, ChanYu2022}. These papers, however, place the prior on
the structural coefficients, which changes the model. The implied prior on $\Pi$ is correlated across equations and depends on the variable ordering, and a reduced-form Minnesota prior can only be approximated \citep{Chan2022}. We instead keep the reduced-form prior of \citet{CarrieroChanClarkMarcellino2022} and use the structural form only as a change of coordinates. The cross-equation coupling then sits in the prior, where it is $\bigO(1)$ against the $\bigO(T)$ information in the likelihood, rather than in the likelihood as in the triangular algorithm
(Remark~\ref{rem:comparison_CTA}).

A third strand, closest to our second contribution, applies sequential Monte
Carlo (SMC) methods to VAR models: \citet{BognanniHerbst2018} do so for
Markov-switching VARs. \citet{BognanniZito2020} develop a sequential estimation
algorithm for a Cholesky-SV VAR in which all time-varying quantities, including
the contemporaneous coefficients, are latent states. The static parameters can
then be marginalised analytically, so their particles need only track state
trajectories, with a mixture-based Gibbs sampler as mutation kernel. Our
setting differs in that $A$ is a static parameter and the Minnesota prior is
non-conjugate, so no such marginalisation is available. Our \smcsq{} sampler instead places the particles on $(A,\Pi,\Phi)$, replaces the intractable likelihood increments with local particle-filter estimates, whose running products are unbiased, and rejuvenates via the exact Particle Gibbs kernel of Section \ref{sec:ourMCMC}. It moreover delivers, at every $t$, both the one-step-ahead predictive density that we use for on-line forecasting and the marginal likelihood that we use for model choice. The latter is not pursued by \citet{BognanniZito2020}, whose focus is the precision and speed of posterior estimation.

\subsection{Plan and notation} 

The paper is organised as follows. Section \ref{sec:model} introduces the model, the reparametrisation of the system on which our sampler rests, and the prior. 

Section \ref{sec:pmcmc} gives a light introduction to the PMCMC (particle MCMC) framework, designed to help the reader better understand how we can use CSMC to sample the volatility trajectories given the parameters. Then, it shows that, under the reparametrisation, the $K$ volatility trajectories are conditionally independent and may each be updated by CSMC. Section \ref{sec:ourMCMC} develops the proposed MCMC sampler for parameter inference. Section \ref{sec:SMC2} develops the proposed SMC$^2$ sampler for on-line prediction and Bayesian model choice, together with its initialisation on a training sample, the equation-by-equation assimilation that avoids imputation on partially observed dates, and the devices that keep its memory cost low. Section \ref{sec:num} contains numerical experiments on simulated and real data that showcase the performance of both algorithms. Section \ref{sec:conclusion} concludes. Appendix \ref{app:cta} compares our kernel with the corrected triangular algorithm. The remaining appendices collect the prior details, the PMMH sampler and the classical Gibbs updates.

We now introduce some of the notation used in the paper. Additional notation will be introduced later in the manuscript. For every integer $M\in\mathbb{N}$,
we let $[M]\coloneq \{1,\ldots,M\}$. For a matrix $W$, we denote by $W^\top$
its transpose, $W_{\cdot,j}$ its $j$-th column, $W_{j,\cdot}$ its $j$-th row
(as a row vector), $W_{(j)}$ its $j$-th row (as a column vector),
\textit{i.e.}, $W_{(j)} \coloneq W_{j,\cdot}^\top$, and by $W_{(1:j)}$ the matrix formed by the first $j$ rows of $W$. We denote by $|W|$ the determinant of a square matrix $W$. For a vector $w\in\mathbb{R}^{K}$, $\diag(w)$ denotes the $K\times K$ diagonal matrix with $w$ on its diagonal. $0_K$ and $I_K$ denote the $K$-dimensional zero vector and identity matrix.  We use extensively the index range notation (also known as the colon notation): \textit{e.g.} $y_{1:T}$ is a short-hand for the collection of variables $(y_1, \ldots, y_T)$, and $y_{1:T,k}$ denotes the $k$-th component of this collection. We use the same notation $\pi$ for the prior and posterior distributions of the parameters and the latent states, as well as for their Lebesgue densities. The intended object is always clear from the arguments. We use $f$ for the densities of the observables. Thus $f(y_{1:t}\mid\theta)$ denotes the likelihood of $y_{1:t}$ at the parameter value $\theta$, with the latent states integrated out, and $f(y_{1:t}) = \int f(y_{1:t}\mid\theta)\,\pi(d\theta)$ denotes its marginal with respect to the prior. In Section \ref{sec:pmcmc}, where $\theta$ is held fixed throughout, we follow the particle MCMC literature and write $\theta$ as a subscript instead. There, $p_\theta(\cdot)$ stands for $f(\cdot\mid\theta)$. We also write $p_\theta(x_{1:t}, y_{1:t})$ for the joint density of the states and the observations, $p_\theta(x_t\mid x_{t-1})$ for the transition density, and $p_\theta(x_{1:t}\mid y_{1:t})$ for the smoothing density, all at the parameter value $\theta$.

\section{The Model}\label{sec:model}

\subsection{Original formulation}

Consider a $K$-dimensional reduced-form VAR($p$) model for $y_t\in\mathbb{R}^K$ with stochastic volatility: for each $t\in[T]$,
\begin{align}
  y_t & = \Pi z_{t} + v_t, \label{model:eq:1}\\
  v_t & = A^{-1}\Lambda_t^{1/2} \epsilon_t, \qquad \epsilon_t
  \stackrel{\text{i.i.d.}}{\sim} \mathcal{N}(0_K,I_K),\label{model:eq:2}
\end{align}
where $z_t \coloneq (1, y_{t-1}^\top, \ldots, y_{t-p}^\top)^\top \in\mathbb{R}^{Kp + 1}$ is a $(Kp + 1)$-vector containing the
first $p$ lags of $y_t$ and the constant, $\Pi \coloneq [\Pi_0,\Pi_1,\ldots,\Pi_p]\in\mathbb{R}^{K\times (Kp + 1)}$ is a $(K\times (Kp+1))$-matrix of coefficients, with $\Pi_0 \in \mathbb{R}^{K}$ the intercept and $\Pi_j \in \mathbb{R}^{K\times K}$ the coefficient matrix of lag $j$, $\Lambda_t \coloneq \diag(\lambda_t)\in\mathbb{R}^{K\times K}$ is a diagonal $K\times K$ matrix whose \(K\)-dimensional vector \(\lambda_t \coloneq (\lambda_{t,1}, \ldots, \lambda_{t,K})^\top\) has the generic $k$-th element \(\lambda_{t,k}\), $A^{-1}\in\mathbb{R}^{K\times K}$ is a lower triangular matrix with ones on its main diagonal, $v_t$ is the $K$-vector of reduced-form errors and $\epsilon_t$ is the $K$-vector of standardized orthogonal innovations. We assume that the VAR($p$) is stable, that is, all roots of the equation $\det(I_K - \Pi_1 u - \Pi_2 u^2 - \ldots - \Pi_p u^p) = 0$ have $|u| > 1$. Stability ensures non-explosive conditional-mean dynamics but does not imply covariance stationarity of $y_t$: under the random-walk log-volatilities of \eqref{eq:same_motion_model}, the unconditional variance of $v_t$ diverges. This combination (stable autoregressive dynamics and nonstationary volatility) is standard in the literature \citep{COGLEY2005, Primiceri2005}.

This specification implies a time-varying covariance matrix $\Sigma_t$ for the error term $v_t$:
\begin{equation*}
  \Sigma_t \coloneq \var(v_t|\lambda_t) = A^{-1}\Lambda_t \left(A^{-1}\right)^\top.
\end{equation*}
Because $\Lambda_t$ is diagonal, the rescaled VAR disturbances $\wtl{v}_t \coloneq A v_t$ have generic element $\wtl{v}_{t,k} = \lambda_{t,k}^{1/2}\epsilon_{t,k}$ mutually independent conditional on $\lambda_t$ and independent across $t$ conditional on $\lambda_{1:T}$. By taking the log, we have:
$$\log \wtl{v}_{t,k}^2 = \log\lambda_{t,k} + \log\epsilon_{t,k}^2,\qquad k\in[K].$$

The law of motion for $\lambda_{t,k}$ is specified as follows. For every $k\in[K]$, the process is initialised at $t=0$, independently across $k$, via
\begin{equation}\label{eq:lambda0_prior}
  \log(\lambda_{0,k}) \sim \mathcal{N}(0, \sigma_0^2),
\end{equation}
and then evolves for $t \in [T]$ according to
\begin{equation}\label{eq:same_motion_model}
  \log(\lambda_{t,k}) = \log(\lambda_{t-1,k}) + e_{t,k}, \qquad e_{t} \stackrel{\text{i.i.d.}}{\sim}  \mathcal{N}(0,\Phi),
\end{equation}
where $e_t \coloneq (e_{t,1},\ldots, e_{t,K})^\top$ and $\Phi\in\mathbb{R}^{K\times K}$ is a diagonal $K\times K$ matrix as in \citet{COGLEY2005}. The process $\{e_t\}_{t\in[T]}$ is independent of $\lambda_0$ and of $\{\epsilon_t\}_{t\in[T]}$. Here $\lambda_{0,k}$ is a latent initial condition: it is not paired with any observation, so the likelihood only depends on $\lambda_{1:T}$. Marginalising over the transition $e_{1,k}$, one obtains $\log(\lambda_{1,k})\sim\mathcal{N}(0, \sigma_0^2+\Phi_{k,k})$. The value of $\sigma_0^2$ is specified in section \ref{sec:prior} below (see \eqref{eq:sigma0_value}).

Given the vector of initial conditions $\by_0 \coloneq (y_0^\top,\ldots,y_{-p+1}^\top)^\top\in\mathbb{R}^{K p}$ and the data $y_{1:T}$, the likelihood of the parameters $(A,\Pi,\lambda_{1:T},\Phi)$ is
\begin{multline}
  f(y_{1:T}|A, \Pi, \lambda_{1:T},\Phi, \by_{0}) = f(y_{1:T}|A, \Pi, \lambda_{1:T},\by_{0})\\ 
  = (2\pi)^{-KT/2}\left\{\prod_{t=1}^T\prod_{k=1}^K\lambda_{t,k}^{-1/2}\right\}
  \exp\left\{-\sum_{k=1}^K\sum_{t=1}^T \frac{(A_{(k)}^\top y_t -
  \wtl{\Pi}_{(k)}^\top z_t)^2}{2\lambda_{t,k}}\right\},
\end{multline}
where $\wtl{\Pi}\coloneq A \Pi$. 
Note that, since $A$ is lower unitriangular, its determinant is $|A| = 1$,
which is why it does not appear above.

\subsection{Re-parameterized model and transformed data}\label{sub:transformed_model}

We re-parameterize the observation  model \eqref{model:eq:1}-\eqref{model:eq:2} as follows: 
\begin{equation}\label{model:eq:reparametrized}
  \wtl{y}_t  = \wtl\Pi z_t + \Lambda_t^{1/2}\epsilon_t, \qquad \epsilon_t \stackrel{i.i.d}{\sim} \mathcal{N}\left(0_K, I_K\right),
\end{equation}
where $\wtl{y}_t \coloneq A y_t$ for $t\in[T]$ and $\wtl\Pi \coloneq A\Pi$, while the law of motion for $\lambda_{1:T}$ remains as in \eqref{eq:lambda0_prior} and \eqref{eq:same_motion_model}.

The likelihood function for the transformed parameters $(A, \wtl\Pi,\lambda_{1:T}, \Phi)$ given the transformed data $\wtl{y}_{1:T}$ and the initial conditions $\mathbf{y}_0$ is: 
\begin{multline}
  f\left(\wtl{y}_{1:T}|A,\wtl{\Pi},\lambda_{1:T},\Phi, \by_{0}\right) = f\left(\wtl{y}_{1:T}|A,\wtl{\Pi},\lambda_{1:T},\by_{0}\right) \\
  = (2\pi)^{-KT/2}\left\{\prod_{t=1}^T|\Lambda_t|^{-1/2}\right\} \exp\left\{-\frac{1}{2}\sum_{t=1}^T (\wtl{y}_t - \wtl{\Pi}z_t)^\top \Lambda_t^{-1}(\wtl{y}_t - \wtl{\Pi}z_{t})\right\}.
\end{multline}

In this paper, we leverage this re-parametrisation to derive our MCMC sampler. That is, we sample from conditional distributions derived from either the original posterior distribution, \textit{i.e.} the distribution of $(A, \Pi, \lambda_{1:T}, \Phi)$ given the data, or the distribution of $(A, \wtl{\Pi}, \lambda_{1:T}, \Phi)$. Sampling from the latter makes it possible to exploit the conditional independence of the components of $\wtl{y}_t$. We return to this point in Section \ref{sec:ourMCMC}.

\subsection{Prior specification}\label{sec:prior}

We assume that the components $A$, $\Pi$, $\Phi$, and $\lambda_0$ are a priori independent, and the prior factorizes: $\pi(A,\Pi,\Phi,\lambda_{0:T}) = \pi(A)\pi(\Pi)\pi(\Phi)\pi(\lambda_0)\prod_{t=1}^T\pi(\lambda_t|\lambda_{t - 1},\Phi)$. Recall that $A$ is lower-triangular, with ones on its diagonal. We specify a prior distribution for $A$ such that the entries below the diagonal are independent and $\mathcal{N}(0, 10^6)$ distributed. This prior is effectively flat over the relevant range: with standardised data, the likelihood concentrates on values of the free elements of $A$ (that is, the coefficients of contemporaneous regressions among the orthogonalised residuals) of at most a few units in magnitude, and over any such range a $\mathcal{N}(0, 10^6)$ density is essentially constant, so the prior exerts no practical influence on the posterior. This is the
almost flat specification of \citet{CarrieroChanClarkMarcellino2022}.

We take the Minnesota prior on $\Pi$ from the same source. What matters for our
purposes is that the VAR coefficients are assigned mutually independent
Gaussian priors, so that the prior covariance of $\operatorname{vec}(\Pi)$ is
diagonal. The means, variances, and shrinkage hyperparameters $\rho \coloneq (\rho_1,\rho_2,\rho_3)^\top$ are the standard Minnesota choices. We record them, together with the intuition behind them, in Appendix \ref{app:minnesota}.

The resulting prior is asymmetric across equations for two reasons: the cross-variable shrinkage $\rho_2 = 0.5 \neq 1$ down-weights other variables' lags relative to own lags, and the variance ratios $\sigma_k^2/\sigma_i^2$ (defined in Appendix \ref{app:minnesota}) scale the shrinkage by equation-specific residual variances. Either feature alone already implies that the diagonal prior covariance of $vec(\Pi)$ cannot be written as a single Kronecker product $\Upsilon\otimes\Omega$, so it lies
outside the natural-conjugate family. 

Stacking the coefficients of equation $j$ into $\Pi_{(j)}\in \mathbb{R}^{Kp + 1}$, the prior is multivariate normal: $\Pi_{(j)} \sim \mathcal{N}_{Kp + 1}(\mu_{\Pi_{(j)}}, \Sigma_{\Pi_{(j)}})$, where $\mu_{\Pi_{(j)}}$ and $\Sigma_{\Pi_{(j)}}$ are derived from the Minnesota hyperparameters of
Appendix \ref{app:minnesota}, so the joint prior decomposes as a sum of per-equation quadratics.\\ 
\indent Using the reparameterization in \eqref{model:eq:reparametrized}, we derive the prior for the $j$-th row of $\wtl{\Pi}$, which is defined as: $\wtl{\Pi}_{(j)}^\top = A_{(j)}^\top \Pi = \sum_{k=1}^j A_{j,k} \Pi_{(k)}^\top$. The Gibbs sampler of Section \ref{sec:ourMCMC} updates $\wtl{\Pi}$ one row at a time, which it can do because the transformed likelihood~\eqref{model:eq:reparametrized} factorises across equations. Each such update requires the full conditional of a single row $\wtl{\Pi}_{(j)}$, whose prior part is $\pi(\wtl{\Pi}_{(j)}\mid\wtl{\Pi}_{(-j)},A)$. This conditional is needed because $A$ makes the rows of $\wtl{\Pi}$ a priori dependent.

Writing $L \coloneq A^{-1}$ (unit lower-triangular) and substituting $\Pi = L\wtl{\Pi}$ into the joint Gaussian prior, one isolates the terms in which $\wtl{\Pi}_{(j)}$ appears and completes the square. This yields a Gaussian conditional prior $\wtl{\Pi}_{(j)} \mid \wtl{\Pi}_{(-j)}, A \sim \mathcal{N}_{Kp + 1}(\mu_{\text{prior}}, \Omega_{\text{prior}})$. The precision is the sum of the precision from equation $j$ and the weighted precisions $L_{kj}^2\,\Sigma_{\Pi_{(k)}}^{-1}$ from all subsequent equations $k > j$:
\begin{equation}\label{eq:prior:precision}
    \Omega_{\text{prior}}^{-1} = \Sigma_{\Pi_{(j)}}^{-1} + \sum_{k=j+1}^{K} L_{kj}^2 \Sigma_{\Pi_{(k)}}^{-1},
\end{equation}
while the mean is
\begin{equation}\label{eq:prior:mean}
    \mu_{\text{prior}} = \Omega_{\text{prior}} \left(\Sigma_{\Pi_{(j)}}^{-1} \mathbf{r}_{j, -j} + \sum_{k=j+1}^{K} L_{kj} \Sigma_{\Pi_{(k)}}^{-1} \mathbf{r}_{k, -j} \right),
\end{equation}
where $\mathbf{r}_{k, -j} \coloneq \mu_{\Pi_{(k)}} - \sum_{m \leq k, m \neq j} L_{km} \wtl{\Pi}_{(m)}$ is the partial prior residual for equation $k$, that is, the part of the prior mean of $\Pi_{(k)}$ not explained by the transformed rows other than the $j$-th, which isolates the prior mean component independent of $\wtl{\Pi}_{(j)}$ and so avoids double-counting. The full completing-the-square derivation is given in Appendix~\ref{app:priorderiv}.\\

Regarding $\Phi$, which is a diagonal matrix, we assume that each component is independently distributed according to an inverse-Gamma prior: $\Phi_{k,k} \sim \mathcal{I}\Gamma\left((K+2)/2,1/2\right)$, for $k\in [K]$. This prior has mean $1/K$ and is tighter than the $\mathcal{IW}((K+2)I_K, K+2)$ of \citet[see Supplement]{CARRIERO2019137}, whose diagonal marginals have mean $K + 2$. 
A tighter prior stabilizes the behavior of the early-stage $SMC^2$. Indeed, in the early stages of $SMC^2$ (when we only have a few observations), the data likelihood is not strong enough to dominate the prior. If we use a loose prior with a large mean like $K+2$, the sampled values for $\Phi_{k,k}$ will be very large. Consequently, the proposed particles for the log-volatility states will spread out, causing the likelihood estimator to have high variance. A tighter prior prevents this and stabilizes the early steps for $SMC^2$, where particles are reweighted by likelihood evaluations. 

Finally, we specify the prior for the initial latent volatility state $(\lambda_{0,k})_{k=1:K}$ introduced in \eqref{eq:lambda0_prior}. The $\lambda_{0,k}$ are mutually independent and, for every $k\in[K]$, we set
\begin{equation}\label{eq:sigma0_value}
  \log(\lambda_{0,k}) \sim \mathcal{N}(0,\, \sigma_0^2), \qquad \sigma_0^2 = 2.31.
\end{equation}
Under \eqref{eq:sigma0_value}, the implied $95\%$ prior interval for the initial variance $\lambda_{0,k}$ is $\exp\{\pm 1.96\sqrt{2.31}\} \approx [0.05,20]$. An interval of this width is informative rather than diffuse: under a much larger value of $\sigma_0^2$, the corresponding interval would span many orders of magnitude. Since the marginal prior variance of $\log(\lambda_{t,k})$ is $\sigma_0^2 + t\Phi_{k,k}$, this shrinkage toward unit error variances is strongest at the start of the sample and decays as increment uncertainty accumulates, which is an unavoidable feature of the random-walk specification. The role of $\sigma_0^2$ is thus to anchor the level from which each volatility path departs. A diffuse initialisation abandons this anchor entirely, leaving the scale of the path essentially unrestricted at every $t$. The anchor matters because $\lambda_{0,k}$ is paired with no observation and early-sample volatilities are the most prior-sensitive part of the path.\\
\indent This prior differs from that of \citet{Primiceri2005}, who centres the initial log-volatilities at training-sample estimates, and from \citet{CarrieroChanClarkMarcellino2022}, who use the same zero-centred Gaussian initialisation but with a much larger variance ($\sigma_0^2 = 100$). Our tighter choice makes the unit-variance centering informative. A diffuse initialisation would also inflate the variance of the particle-filter estimates of the likelihood that underlie the sequential algorithm of Section \ref{sec:pmcmc}, since initial particles drawn from a diffuse prior mostly land at implausible volatility scales. Recall that $\lambda_{0,k}$ is a latent initial condition. It does not enter the likelihood, which depends only on $\lambda_{1:T,k}$. Its prior still matters. Under the random walk, the implied prior on each $\lambda_{t,k}$ is centred at the prior mean of $\lambda_{0,k}$.

\section{Particle MCMC}\label{sec:pmcmc}

\subsection{A quick introduction to PMCMC}\label{sec:introduction:PMCMC}

For the sake of exposition, we present the general PMCMC methodology of \citet{PMCMC} (introduced to the econometric literature by \citet{FluryShephard2011}, see \citet{Creal2012} for a survey of sequential Monte Carlo (SMC) methods in economics and finance) in the context of a basic univariate stochastic volatility model, \textit{i.e.},
\begin{align}\label{eq:basicSV}
  y_t & = \lambda_t^{1/2} \epsilon_t, \qquad \epsilon_t \stackrel{i.i.d.}{\sim} \mathcal{N}(0, 1),\qquad t\in[T], \nonumber\\
  \log (\lambda_t) & = \log(\lambda_{t-1})+ e_t,\qquad e_t \stackrel{i.i.d.}{\sim} \mathcal{N}(0,\sigma^2),\qquad t\in[T],
\end{align}
where $\log(\lambda_0)\sim \mathcal{N}(0, \sigma_0^2)$ is a latent initial condition that is not paired with any observation, and $\{\epsilon_t\}_{t\in [T]}$, $\{e_t\}_{t\in [T]}$, and $\lambda_0$ are mutually independent. Here the parameter is $\theta=\sigma^2$, and $\sigma_0^2$ is a fixed prior variance.

Such a model is an instance of a state-space model (also known as a hidden
Markov model), in which the observations depend on an unobserved Markov
process. Here the unobserved process is $x_t \coloneq \log \lambda_t$, and it follows a Gaussian random walk. This model is a workhorse in financial econometrics (\textit{e.g.}, \citet{KimShephardChib1998}) and is often estimated via MCMC methods that rely on mixture approximations.

In contrast, we use particle filters to approximate the filtering distribution $\pi(x_t|y_{1:t}, \theta)$ recursively through a sequential Monte Carlo algorithm that propagates, reweights, and resamples $N$ particles $x_t^n$ (candidate values) recursively. 
Particle filters are by now a standard tool in macroeconometrics for the likelihood evaluation of nonlinear state-space models \citep{FernandezVillaverdeRubioRamirez2007, HerbstSchorfheide2016}. Algorithm \ref{alg:bootstrap} describes the simplest particle filtering algorithm for such a model, which is often referred to as the bootstrap particle filter \citep{Gordon}. There, particles are simulated according to the dynamics of
the model, reweighted according to the likelihood of data point $y_t$, given $x_t$, and then resampled.

\begin{algorithm}
  \caption{Bootstrap filter for basic univariate SV model}
  \label{alg:bootstrap}
  \begin{algorithmic}
    \Require data $y_{1:T}$, parameter $\theta = \sigma^2$, prior variance $\sigma_0^2$
    \State $x_0^n \sim \mathcal{N}(0, \sigma_0^2)$ for $n\in[N]$
    \hfill\Comment{Draw initial latent state (no observation at $t=0$)}
    \For{$t=1,\dots,T$}
      \If{$t>1$}
        \State $a_t^{1:N} \sim \text{Cat}(W_{t-1}^{1:N})$
        \hfill\Comment{Resampling}
      \Else
        \State $a_1^n \gets n$ for $n\in[N]$ \hfill\Comment{No resampling at $t=1$}
      \EndIf
      \State $x_t^n \sim \mathcal{N}(x_{t-1}^{a_t^n}, \sigma^2)$ for $n\in[N]$
      \hfill\Comment{Propagate via transition}
      \State $w_t^n \gets p_\theta(y_t|x_t^n) = \varphi(y_t; 0, e^{x_t^n})$, for $n\in[N]$
      \State $W_t^n \gets w_t^n / \sum_{m=1}^N w_t^m$, for $n\in[N]$
    \EndFor
  \end{algorithmic}
  \medskip
  \footnotesize\emph{Note:} $\text{Cat}(W_{t-1}^{1:N})$ denotes the categorical distribution which returns $n$ with probability $W_{t-1}^n$, for $n\in [N]$, and $W_{t-1}^{1:N}\coloneq (W_{t-1}^1, \dots, W_{t-1}^N)$. $a_t^{1:N}\sim \mathrm{Cat}(W_{t-1}^{1:N})$ means $N$ i.i.d. draws.
  $\varphi(y;\mu,s^2)$ denotes the $\mathcal{N}(\mu,s^2)$ density evaluated at $y$.
\end{algorithm}


Beyond filtering, particle filters such as Algorithm \ref{alg:bootstrap} also provide an estimate of the likelihood. At any time $t$, the quantity
\begin{equation*}
  \wh p_\theta(y_{1:t}) \coloneq \prod_{s=1}^t \left(\frac 1 N \sum_{n=1}^N w_s^n\right)
\end{equation*}
is an unbiased estimate of $p_\theta(y_{1:t}) = \int p_\theta(x_{1:t}, y_{1:t}) dx_{1:t}$ \citep{DelMoral1996unbiased}. This unbiasedness is the key property exploited by particle MCMC, to which we now turn.

\citet{PMCMC} proposed a general approach, known as particle MCMC (PMCMC), to sampling from the posterior distribution of the parameter $\theta$ of a state-space model. In a nutshell, the approach amounts to designing MCMC kernels that move around the $\theta$-space and, for each newly proposed value of $\theta$, run a new particle filter at that value, and use its output (the likelihood estimate above, which replaces the true likelihood, or a particle trajectory, which serves as a draw of the latent states) in an otherwise standard MCMC step. Remarkably, the resulting kernels leave the exact posterior invariant for any fixed number of particles $N$. The choice of $N$ affects only the mixing of the chain, not its target. 

As a first example of such a kernel, PMMH (particle marginal Metropolis-Hastings) is a Metropolis sampler that proposes, at each MCMC iteration, a new $\theta$-value (say, according to a random walk proposal) and runs a particle filter at this value. The Metropolis acceptance probability involves the likelihood, which is intractable here. PMMH replaces it with the particle filter estimate. Since this estimate is unbiased, the resulting chain still targets the exact posterior.

For $N$ large enough, we expect PMMH to behave essentially like a standard random walk Metropolis sampler in which the likelihood is computed exactly. Exactness, however, does not require a large $N$. As stated above, for any fixed $N$ the PMMH kernel leaves the true posterior distribution of $\theta$ invariant. Taking $N$ small only degrades the mixing of the chain. We refer to Appendix \ref{app:pmmh} (Algorithm \ref{alg:pmmh}) for a description of PMMH, and to \citet{PMCMC} and \citet[][Chap.~16]{SMCbook} for more background on this approach.

PMMH targets the parameter $\theta$. In the setting \eqref{eq:basicSV}, however, the sampler must also draw the full volatility path given the parameters. This is the purpose of CSMC (conditional sequential Monte Carlo), a second PMCMC kernel, introduced in \citet{PMCMC} under the name conditional SMC. A CSMC step is one iteration of an MCMC algorithm on the space of state trajectories, and it targets the smoothing distribution $p_\theta(x_{1:T} \mid y_{1:T})$. The kernel takes as input a state trajectory $x_{1:T}$ and runs a particle filter, such as Algorithm \ref{alg:bootstrap}, conditionally on one particle path coinciding with the input. Namely, at every time $t$, particle $1$ is set to $x_t^1 = x_t$ and its ancestor to $a_t^1 = 1$, while the remaining $N-1$ particles are propagated, reweighted and resampled as usual. Once the particle filter has been run, a new trajectory $(x_1^{b_1}, \dots, x_T^{b_T})$ is sampled from the set of $N^T$ potential trajectories by drawing the indices $b_{1:T}$ in a backward pass. For any fixed $N \geq 2$, the CSMC kernel leaves $p_\theta(x_{1:T} \mid y_{1:T})$ invariant. Algorithm \ref{alg:CSMC} describes the CSMC kernel that corresponds to the univariate stochastic volatility model.

\begin{algorithm}
  \caption{CSMC (conditional Sequential Monte Carlo) kernel with backward sampling step for the basic univariate SV model}
  \label{alg:CSMC}
  \begin{algorithmic}
    \Require input state trajectory $x_{1:T}$, parameter $\theta = \sigma^2$, prior variance $\sigma_0^2$
    \State $x_0^n \sim \mathcal{N}(0, \sigma_0^2)$ for $n\in[N]$
    \hfill\Comment{Draw initial latent state (not conditioned, no observation at $t=0$)}
    \For{$t=1,\dots,T$} 
    \hfill \Comment{Forward pass}
      \If{$t>1$}
        \State $a_t^1 \gets 1$ 
        \State $a_t^{2:N}\sim \text{Cat}(W_{t-1}^{1:N})$ 
        \hfill \Comment{Resampling}
      \Else
        \State $a_1^n \gets n$ for $n\in[N]$ \hfill\Comment{No resampling at $t=1$}
      \EndIf
      \State $x_t^1 \gets x_t$
      \hfill\Comment{Force particle 1 onto the reference trajectory}
      \State $x_t^n \sim \mathcal{N}(x_{t-1}^{a_t^n}, \sigma^2)$ for $n=2,\dots, N$
      \hfill\Comment{Propagate via transition}
      \State $w_t^n \gets p_\theta(y_t|x_t^n) = \varphi\left(y_t; 0, e^{x_t^n} \right)$, for $n\in[N]$
      \State $W_t^n \gets w_t^n / \sum_{m=1}^N w_t^m$, for $n\in[N]$
    \EndFor
    \State $b_T\sim \text{Cat}(W_T^{1:N})$
    \For{$t=T-1,\dots, 1$} 
      \Comment{Backward pass: pick one particle at each  time $t$}
      \State $\widehat{w}_t^n \gets W_t^n \times p_\theta(x_{t+1}^{b_{t+1}}|x_t^n)$ 
        for $n\in[N]$
      \State $\widehat{W}_t^n = \widehat{w}_t^n / \sum_{m=1}^N \widehat{w}_t^m$
      \State $b_t\sim \text{Cat}(\widehat{W}_t^{1:N})$ 
      \EndFor
      \State 
    \Return $(x_1^{b_1}, \dots, x_T^{b_T})$ 
    \hfill\Comment{Output trajectory $x_{1:T}$}
  \end{algorithmic}
  \medskip
\footnotesize\emph{Note:} Notation as in Algorithm \ref{alg:bootstrap}. The kernel takes a trajectory $x_{1:T}$ as input and returns a new trajectory
$(x_1^{b_1},\dots,x_T^{b_T})$. Iterated over MCMC sweeps, it leaves $p_\theta(x_{1:T}\mid y_{1:T})$ invariant. In the backward pass, $p_\theta(x_{t+1}\mid x_t)$ denotes the transition density, here $\varphi(x_{t+1};\, x_t,\, \sigma^2)$. The backward sampling step \citep{Whiteley_disc_PMCMC} is optional, see the discussion in the text.
\end{algorithm}

Note that Algorithm \ref{alg:CSMC} includes a backward sampling step, due to \citet{Whiteley_disc_PMCMC}, which redraws the index $b_t$ at every time step. This step is optional. A simpler alternative draws $b_T \sim \mathrm{Cat}(W_T^{1:N})$ as in Algorithm \ref{alg:CSMC}, and then replaces the backward pass by ancestral tracing, that is, it `pulls' the history of particle $b_T$ by setting $b_t \gets a_t^{b_{t+1}}$ for $t = T-1, \dots, 1$. Backward sampling, however, has been shown to greatly improve mixing, especially when $T$ is large (\citet{lee2020coupled}). We emphasise the contrast with the standard econometric treatment of stochastic volatility, which replaces the $\log\chi^2_1$ distribution of $\log \epsilon_t^2$ with a discrete mixture of normals (\citet{KimShephardChib1998}): PMCMC targets the exact posterior for any fixed number of particles $N \geq 2$, with no mixture approximation error and no need for the associated ordering correction of \citet{DelNegroPrimiceri2015}.

\begin{figure}[htpb]
    \centering
    \includegraphics[width=0.9\linewidth]{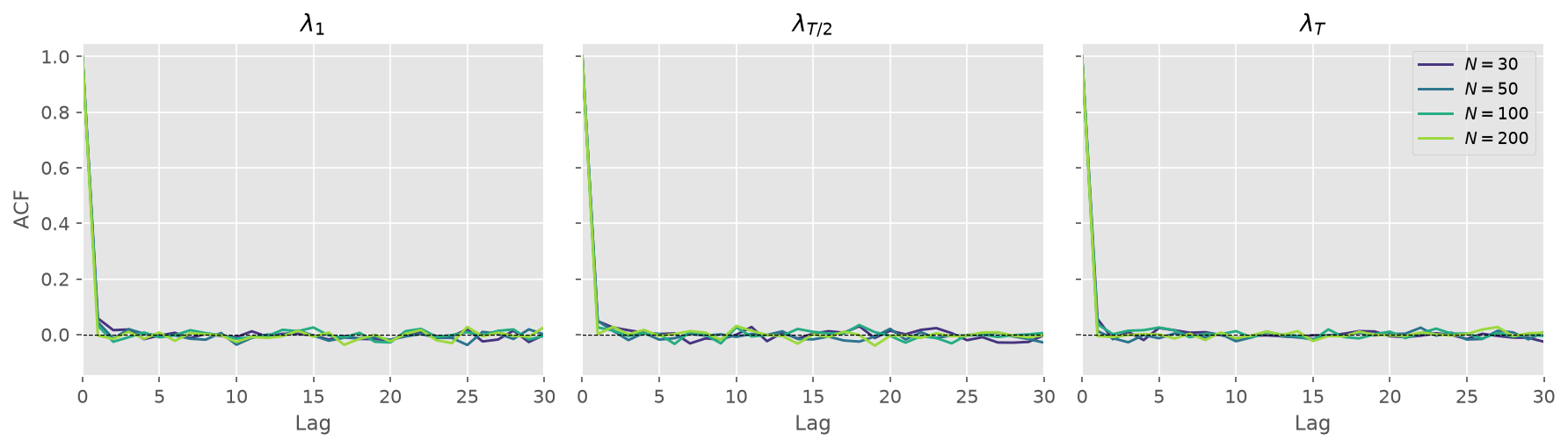}
    \caption{ACF from CSMC algorithm (Algorithm~\ref{alg:CSMC}) for latent variable $\lambda_t$ at the beginning, middle, and end of the trajectory. For various number of particles $N$, each chain is run for $5,000$ iterations after a $500$-step burn-in.}
    \label{fig:csmc_acf}
\end{figure}

To explicitly assess the mixing of Algorithm~\ref{alg:CSMC}, we conduct a numerical experiment with simulated data from the univariate stochastic volatility model in equation \eqref{eq:basicSV} ($T=300$, initial state variance $\sigma_0^2 = 1.0$, and innovation variance $\Phi =
0.05$). By fixing the static parameters to their true values used in data generation, we isolate the performance of the CSMC update step. The algorithm
is run across various number of particles $N$. Figure~\ref{fig:csmc_acf} showcases the ACF (auto-correlation function). We observe a rapid decay of ACF
at every presented time step even for low values of $N$. In particular, the rapid decay at early time steps $\lambda_1$ demonstrates the effectiveness of
the backward sampling step in mitigating path degeneracy.

\subsection{Relevance for our model}\label{sec:relevance:our:model}
We could in principle implement PMMH for our model \eqref{model:eq:1} and \eqref{model:eq:2}. However, this approach has several limitations. First, the dimension of the parameter $\theta = (A, \Pi, \Phi)$ is $\bigO(K^2p)$, which is potentially very large. Random walk Metropolis scales poorly with the dimension: to maintain a reasonable acceptance rate, the standard deviation of the proposal increments must shrink as the dimension grows, so that the number of iterations needed to explore the parameter space grows linearly in the parameter dimension. Second, calibrating the proposal, that is, choosing its covariance matrix for good mixing, is notoriously difficult in the context of PMMH, and even more so when the dimension is large, see the discussion in \citet[Sect.~16.5]{SMCbook}. Third, the acceptance rate of PMMH deteriorates as the variance of the log-likelihood estimate increases. Keeping this variance roughly constant requires the number of particles $N$ to grow with both the sample size $T$ and the dimension of the latent state \citep{Pitt2012134,MR3371005,Sherlock2015}, see Appendix \ref{app:pmmh}. In our setting the latent state is the $K$-dimensional vector $\log\lambda_t$ and the sample counts several hundred observations, so PMMH is prohibitively expensive.

Instead, in this paper we use the CSMC approach and apply it to the reparametrised model \eqref{model:eq:reparametrized}. The key point is that, under the reparametrised model, the $K$ volatility trajectories are conditionally independent given the parameters and the data. Namely,
\begin{equation}\label{eq:lambda_factorisation}
  \pi\left(\lambda_{1:T} \left| A, \wtl{\Pi}, \Phi, \wtl{y}_{1:T}, \by_0\right.\right)
  = \prod_{k=1}^K \pi\left(\lambda_{1:T,k} \left| A, \wtl{\Pi}, \Phi, \wtl{y}_{1:T}, \by_0\right.\right),
\end{equation}
and the $k$-th factor is the smoothing distribution of a univariate state-space
model. Its hidden process is $x_{t,k} = \log \lambda_{t,k}$, the Gaussian
random walk \eqref{eq:same_motion_model} with innovation variance $\Phi_{k,k}$.
Its observations are the residuals $r_{t,k} \coloneq \wtl{y}_{t,k} -
\wtl{\Pi}_{(k)}^\top z_t$, with $r_{t,k} \mid x_{t,k} \sim \mathcal{N}(0,
e^{x_{t,k}})$. In terms of $x_{1:T,k} \coloneq \log\lambda_{1:T,k}$, this smoothing density is explicit up to its normalising constant. Marginalising the initial condition $x_{0,k}$ gives $x_{1,k} \sim \mathcal{N}(0, \sigma_0^2 + \Phi_{k,k})$. By Bayes' theorem, the smoothing density is proportional to the prior on the path times the likelihood:
\begin{multline}
  \pi\left(x_{1:T,k} \mid A, \wtl{\Pi}, \Phi, \wtl{y}_{1:T}, \by_0\right)
  \propto \varphi\big(x_{1,k};\, 0,\, \sigma_0^2 + \Phi_{k,k}\big) \prod_{t=2}^{T} \varphi\big(x_{t,k};\, x_{t-1,k},\, \Phi_{k,k}\big)
  \prod_{t=1}^{T} \varphi\big(r_{t,k};\, 0,\, e^{x_{t,k}}\big)\\
  \propto
  \exp\Bigg\{-\frac{x_{1,k}^2}{2(\sigma_0^2 + \Phi_{k,k})} - \sum_{t=2}^{T} \frac{(x_{t,k} - x_{t-1,k})^2}{2\,\Phi_{k,k}} - \frac{1}{2}\sum_{t=1}^{T} \left( x_{t,k} + r_{t,k}^2\, e^{-x_{t,k}} \right)\Bigg\}.\label{eq:smoothing_explicit}
\end{multline}
The normalising constant is the likelihood of the residual series $r_{1:T,k}$ with its state trajectory integrated out, an intractable $T$-dimensional integral. It marginalises the latent path at fixed parameters, unlike the marginal likelihood of Section~\ref{sec:SMC2}, which marginalises the parameters. The observation terms $r_{t,k}^2 e^{-x_{t,k}}$ are not quadratic in $x_{t,k}$, so the model is not linear Gaussian and the Kalman smoother does not apply. The CSMC kernel of Algorithm~\ref{alg:CSMC} leaves \eqref{eq:smoothing_explicit} invariant without approximating it. The factorisation \eqref{eq:lambda_factorisation} holds because both the likelihood and the prior on the volatility paths are products over $k$. The likelihood factorises because $\Lambda_t$ is diagonal. The prior factorises because $\Phi$ is diagonal and the initial conditions $\log\lambda_{0,k}$ are independent across $k$. We can therefore update each trajectory $\lambda_{1:T,k}$ with the CSMC kernel of Algorithm~\ref{alg:CSMC}, applied to the observed series $r_{1:T,k}$. As we have shown in the previous section, this yields excellent mixing for these state trajectories even with small $N$.

\section{Proposed MCMC kernel}\label{sec:ourMCMC}

\subsection{General structure}

We are now ready to describe our MCMC approach. The corresponding kernel has the following Gibbs-like structure:
\begin{enumerate}
  \item $A | \Pi, \Phi, \lambda_{1:T}, y_{1:T}$
  \item $\wtl{\Pi} | A, \Phi, \lambda_{1:T}, \wtl{y}_{1:T}$
  \item $\Phi | A, \wtl{\Pi}, \lambda_{1:T}, \wtl{y}_{1:T}$
  \item $\lambda_{1:T} | A, \wtl{\Pi}, \Phi, \wtl{y}_{1:T}$
\end{enumerate}
In fact, several conditionals depend on subsets of these variables. For example, $\Phi$ enters neither Step 1 nor Step 2, and Step 3 depends on $\lambda_{1:T}$ alone (see Appendices~\ref{app:updateA} and~\ref{app:updatePhi}). Note also that the transformed data $\wtl{y}_t = A y_t$ must be recomputed at the end of Step 1, since Step 1 updates $A$. Since $A$ is invertible and appears in every conditioning set, conditioning on $\wtl{y}_{1:T}$ in Steps 2 to 4 is equivalent to conditioning on $y_{1:T}$. We write $\wtl{y}_{1:T}$ to emphasise that these steps operate on the reparametrised model. For the same reason, the parametrisation in which a step is expressed does not affect its validity. Given $A$, the map $\Pi \mapsto \wtl{\Pi} = A\Pi$ is a bijection, and $A$ belongs to the conditioning set of Step 2. The conditional distribution of $\Pi$ given $(A, \Phi, \lambda_{1:T}, y_{1:T})$ is therefore the image of the conditional distribution of $\wtl{\Pi}$ given the same variables under $w \mapsto A^{-1} w$. Consequently, drawing $\wtl{\Pi}^\star$ from the full conditional of $\wtl{\Pi}$ and setting $\Pi^\star = A^{-1} \wtl{\Pi}^\star$ produces an exact draw of $\Pi$ from its full conditional. Step 2 is thus a standard Gibbs update of $\Pi$, computed in more convenient coordinates.

In summary, Steps 1 to 3 are exact draws from full conditional distributions of the joint posterior. Step 4 is not. The CSMC update does not draw $\lambda_{1:T}$ from its full conditional. Instead, it applies a Markov kernel that leaves this conditional invariant \citep{PMCMC}. Strictly speaking, our sweep is therefore not a Gibbs sampler but a Particle Gibbs sampler in the sense of \citet{PMCMC}. It remains a valid MCMC kernel. The relevant general principle is the following: if each of several Markov kernels leaves a given distribution $\pi$ invariant, then their composition, taken in any order, also leaves $\pi$ invariant \citep{Tierney1994}. In our case, each of Steps 1 to 4 defines a Markov kernel on the space of $(A, \Pi, \Phi, \lambda_{1:T})$, and each of these four kernels leaves the joint posterior invariant. By the principle above, their composition, that is, the full sweep, leaves the joint posterior invariant as well, whatever the ordering of the steps. The ordering subtleties analysed by \citet{DelNegroPrimiceri2015} arise specifically from the mixture-indicator augmentation of \citet{KimShephardChib1998}. In that sampler, some blocks are drawn conditionally on the indicators and others marginally of them, and the validity of such partially collapsed sweeps depends on the order of the steps \citep{vanDykPark2008}. Our sampler involves no indicators, so the issue does not arise.

The MCMC literature has long exploited alternative parametrisations \citep{Papaspiliopoulos2003, Papaspiliopoulos2007, yu2011center, kastner_asis}. The typical setting is a hierarchical model that admits both a centred parametrisation, in which the latent states enter the observation density directly, and an uncentred one, in which the states are rewritten as functions of innovations that are a priori independent of the parameters. Neither parametrisation dominates in general, and alternating between the two leads to better mixing.

Our primary motivation is different. As shown in Section~\ref{sec:relevance:our:model}, the reparametrisation makes the volatility trajectories $\lambda_{1:T,k}$ conditionally independent in Step 4, given the parameters and $\wtl{y}_{1:T}$, see \eqref{eq:lambda_factorisation}. As an added benefit, the reparametrisation also improves the mixing of the update of $\Pi$ (or, rather, of $\wtl{\Pi}$), as we explain in the next section. Interestingly, our reparametrisation is not an uncentred one. It transforms the observations rather than the latent states, and its purpose is to make the $K$ components of the transformed observations conditionally independent, not to disentangle the latent states from the parameters. In this sense it is closer to an orthogonalisation.

Since Steps 1 and 3 are classical Gibbs updates, we defer their derivations to Appendices \ref{app:updateA} and \ref{app:updatePhi} for $A$ and $\Phi$, respectively. Step 4 performs one CSMC step per trajectory $\lambda_{1:T,k}$, $k \in [K]$, as discussed in Section \ref{sec:relevance:our:model}. Since the $K$ trajectories are conditionally independent given $(A, \wtl{\Pi}, \Phi, \wtl{y}_{1:T})$, see \eqref{eq:lambda_factorisation}, these $K$ updates may be
carried out in parallel, which is a further practical advantage of the reparametrisation. The remainder of this section focuses on Step 2.

\subsection{Updating $\wtl{\Pi}$ instead of $\Pi$} 

Recall that matrix $\Pi$ has $d_\pi \coloneq K(Kp + 1) = \bigO(K^2 p)$ entries. It is easy to check that the full conditional posterior distribution of $\Pi$, that is, its posterior distribution conditional on $(A, \lambda_{1:T})$ ($\Phi$ being redundant here), is Gaussian. The standard way to sample from this Gaussian distribution requires computing the Cholesky decomposition of the covariance matrix, leading to a $\bigO(d_\pi^3) = \bigO(K^6 p^3)$ complexity.

We propose instead to update successively the $K$ rows of $\wtl{\Pi}$, that is, $ \wtl{\Pi}_{(1)}, \dots, \wtl{\Pi}_{(K)}$. The likelihood of the reparametrized model actually factorises over the $\wtl{\Pi}_{(k)}$:
\begin{equation}
  f\left(\wtl{y}_{1:T} \left| A, \wtl{\Pi}, \Phi, \lambda_{1:T}, \by_0\right. \right) \propto \prod_{k = 1}^K \prod_{t=1}^T \exp\left\{ - \frac{(\wtl{y}_{t,k} - \wtl{\Pi}_{(k)}^\top z_t)^2}{2\lambda_{t, k}}\right\}.
\end{equation}

On the other hand, the rows of $\wtl{\Pi}$ are not independent a priori, except in the trivial case $A = I_K$. We can still exploit the factorisation above. Namely, we update the rows one at a time, sampling each $\wtl{\Pi}_{(k)}$ from its conditional distribution given the other rows $\wtl{\Pi}_{(k')}$, $k' \neq k$, and the remaining parameters. In this decomposition, the rows are coupled only through the prior, whereas in the original parametrisation they are coupled through the likelihood itself. When the likelihood is informative relative to the prior, this coupling is weak, the rows are nearly conditionally independent, and the equation-by-equation Gibbs update mixes well. Our numerical experiments in Section \ref{sec:num} confirm this behaviour.

More precisely, the full conditional distribution of $\wtl{\Pi}_{(k)}$,
conditional on $\wtl{\Pi}_{(-k)} \coloneq \{\wtl{\Pi}_{(k')}, \,k'\neq k\}$, the other parameters and the data is:
\begin{align*}
  \pi\left(\left.\wtl{\Pi}_{(k)} \right| \wtl{\Pi}_{(-k)}, A, \Phi, \lambda_{1:T}, \wtl{y}_{1:T},\by_0\right) & \propto  f\left(\wtl{y}_{1:T} \left| A, \wtl{\Pi}, \lambda_{1:T}, \by_0 \right.\right) \pi\left(\left.\wtl{\Pi}_{(k)} \right| \wtl{\Pi}_{(-k)}, A\right) \\
   & \propto \exp\left\{ -\sum_{t=1}^T \frac{(\wtl{y}_{t,k} - \wtl{\Pi}_{(k)}^\top z_t)^2}{2\lambda_{t,k}} \right\}
  \pi\left(\left.\wtl{\Pi}_{(k)} \right| \wtl{\Pi}_{(-k)}, A\right),
\end{align*}
and, since both factors are Gaussian (in $\wtl{\Pi}_{(k)}$), the product is also Gaussian:
\begin{equation}
    \pi\left(\left.\wtl{\Pi}_{(k)} \right| \wtl{\Pi}_{(-k)}, A, \Phi, \lambda_{1:T}, \wtl{y}_{1:T}, \by_0\right) = 
  \mathcal{N}\left(\bar{\mu}_{\wtl{\Pi}_{(k)}}, \bar{\Sigma}_{\wtl{\Pi}_{(k)}}\right),
\end{equation}
with
\begin{align}
    \bar{\Sigma}_{\wtl{\Pi}_{(k)}}^{-1} &= \underbrace{\left( \sum_{t=1}^T \frac{z_t z_t^\top}{\lambda_{t,k}} \right)}_{\text{Likelihood}} + \underbrace{\left( \Sigma_{\Pi_{(k)}}^{-1} + \sum_{m = k+1}^{K} L_{m k}^2 \Sigma_{\Pi_{(m)}}^{-1} \right)}_{\text{Corrected Prior Precision}} \\
    \bar{\mu}_{\wtl{\Pi}_{(k)}} & = \bar{\Sigma}_{\wtl{\Pi}_{(k)}} \left[ \sum_{t=1}^T \frac{z_t \wtl{y}_{t,k}}{\lambda_{t,k}} + \left( \Sigma_{\Pi_{(k)}}^{-1} \mathbf{r}_{k, -k} + \sum_{m = k+1}^{K} L_{m k} \Sigma_{\Pi_{(m)}}^{-1} \mathbf{r}_{m, -k} \right) \right].
\end{align}

\begin{remark}\label{rem:comparison_CTA} 
Our update of $\wtl{\Pi}$ is close in spirit to the corrected triangular algorithm (CTA) of \citet{CarrieroChanClarkMarcellino2022}, which amends the
original triangular algorithm of \citet{CARRIERO2019137}. Both are equation-by-equation Gibbs sweeps, and both admit implementations that run at $\bigO(TK^3p^2 + K^4p^3)$ per sweep. They differ in how the likelihood moments are accumulated. Row $j$ of our update involves only its own transformed observations $\wtl{y}_{1:T,j}$, alongside the regressors $z_{1:T}$ that are common to all equations, whereas block $j$ of the CTA augments equation $j$ with all downstream equations $m \geq j$, and so involves a weighted sum of the volatility-weighted Gram matrices of all of them. Forming those $K$ matrices costs $\bigO(TK^3p^2)$ per sweep and is common to both samplers. Once they are cached, which the sums-of-moments form of the corrigendum makes possible, the recombination adds only $\bigO(K^4p^2)$, no more than the $\bigO(K^4p^3)$ that both samplers pay for their $K$ factorisations of $(Kp+1)$-dimensional precisions. The code released with the corrigendum instead rebuilds a stacked design of $T(K-j+1)$ rows for each $j$, raising the accumulation cost to
$\bigO(TK^4p^2)$, an artefact of that implementation rather than of the algorithm (Appendix~\ref{app:cta}).

Since the two updates cost essentially the same, the difference that matters is statistical rather than computational, and it concerns mixing. The reparametrisation $\wtl{\Pi} = A\Pi$ moves the cross-equation coupling out of the likelihood and into the prior. The off-diagonal blocks of the conditional precision are then $\bigO(1)$, while the diagonal blocks are of order $T$, so the partial correlations between rows are $\bigO(1/T)$ and the rows decouple as data accumulate. In the CTA the coupling is likelihood-borne, and the off-diagonal blocks grow with $T$ at the same rate as the diagonal ones, so the partial correlations do not vanish. A full comparison, including the complexity accounting and
the mixing-rate argument, is given in Appendix~\ref{app:cta}.
\end{remark}

\begin{remark}[No data augmentation under partial observations]\label{rem:no_augmentation}
The second advantage is non-asymptotic and matters when the current date is only partially assimilated, as happens within the \smcsq{} recursion of Section~\ref{sec:SMC2}. There, the components of $y_t$ are processed one at a time, so at the sub-step $(t,k)$ the target conditions on $y_{1:t-1}$ and on the first $k$ components $y_{t,1:k}$ only, while the tail $y_{t,k+1:K}$ has not yet been assimilated. Every earlier date has by then been assimilated in full, so the current date is the only incomplete one. Such a date requires no data augmentation in our sweep: the withheld components are neither added to the sampler's state nor simulated at any step. The likelihood factor of row $j$ involves no other equation's coefficients and, crucially, no other equation's contemporaneous observations: the lagged values of all $K$ series enter only through the common regressor $z_t$. Because $A$ is unit lower triangular, the transformed observation $\wtl{y}_{t,j} = A_{(j)}^\top y_t$ of an assimilated equation $j \leq k$ involves only $y_{t,1:j}$ and is unaffected, while the likelihood sums of the equations $j > k$ are simply truncated at $t-1$ (Section~\ref{sub:smc2:rejuv}). The cross-equation couplings sit entirely in the prior, see \eqref{eq:prior:precision} and \eqref{eq:prior:mean}, whose terms involve $A$, the remaining rows $\wtl{\Pi}_{(-j)}$, and the prior hyperparameters, but no data. In \citet{CarrieroChanClarkMarcellino2022}, by contrast, the full conditional of the equation-$j$ coefficients involves the data of all downstream equations $i \geq j$. Each downstream term contains the shock $A_{(i)}^\top v_t$, a linear combination of the contemporaneous residuals $v_{t,1:i}$, which cannot be formed when any component of $y_{t,1:i}$ is missing. Since the downstream set $\{j, \dots, K\}$ always contains unobserved equations, every coefficient block is affected. The remedies are either to augment the sampler with the withheld values, simulated afresh at every sweep, or to remove the incomplete time-$t$ terms separately inside each of the $K$ conditionals. Our sampler needs neither. This property will be convenient for the \smcsq{} sampler developed in the next section.
\end{remark}

\section{On-line prediction and model choice: \smcsq}\label{sec:SMC2}

The MCMC kernel of Section~\ref{sec:ourMCMC} targets the posterior distribution given a fixed sample $y_{1:T}$. We now turn to the sequential setting motivated in the introduction: computing the posterior $\pi(\theta|y_{1:t})$, the predictive distribution of $y_{t+1}$, and the marginal likelihood $f(y_{1:t})$ recursively as observations accrue, so as to perform on-line forecasting and model choice. The sampler we develop for this purpose, \smcsq, requires a rejuvenation kernel that leaves each partial posterior invariant (more precisely, an extended version of it, see Section \ref{sub:smc2:generic}). The kernel of Section \ref{sec:ourMCMC}, and in particular its ability to handle partially observed periods without imputation (Remark \ref{rem:no_augmentation}), is what makes this step feasible in our model.\\
\indent We proceed in two steps. While Section \ref{sec:pmcmc} used SMC to track the latent states for a fixed parameter, the same principles (importance sampling, resampling, MCMC moves) can be applied to the static parameter $\theta$ itself, by targeting the sequence of partial posteriors $\pi_t(\theta) \coloneq \pi(\theta|y_{1:t})$, for $t \geq 1$, with $\pi_0(\theta) \coloneq \pi(\theta)$ denoting the prior distribution of $\theta$. SMC samplers of this type \citep{DelMoral2006} have become a standard tool for posterior simulation in econometrics (\textit{e.g.} \cite{HerbstSchorfheide2014}). We first present IBIS \citep[Iterated batch importance sampling]{MR1929161}, a SMC sampler that makes it possible to perform Bayesian sequential inference, prediction, and model choice for data $y_1, y_2, \dots$, under the assumption that one is able to compute the likelihood $f(y_t|y_{1:t-1}, \theta)$ of each new data-point $y_t$ given the past data and the parameter $\theta$. This quantity is not tractable in our model (or more generally for any state-space model), as it is an integral with respect to the latent states.\\
\indent Thus, in a second step, we will replace IBIS by a more sophisticated algorithm, \smcsq{}, which removes this assumption by replacing the exact increment with a particle-filter estimate, and which we adapt to our setting in Sections \ref{sub:smc2:init} - \ref{sub:smc2:rejuv}. However, presenting IBIS will greatly facilitate the exposition of \smcsq{}.

\subsection{IBIS algorithm}
One may approximate the distributions $\pi_t$ recursively through sequential importance sampling. At time $0$, one samples $\theta^l\sim\pi_0$, $l\in[N_\theta]$, and initialises weights as $w_0^l \gets 1$. Then, at each time $t$, one may update the weights as follows:
$\omega_t^l \gets \omega_{t-1}^l \times u_t(\theta^l)$, so that $\omega_t^l = \prod_{s=1}^t u_s(\theta^l) = f(y_{1:t}|\theta^l)$ by the chain rule, with 
\[
  u_t(\theta) \coloneq f(y_t|y_{1:t-1}, \theta) \propto
  \frac{\pi_t(\theta)}{\pi_{t-1}(\theta)}. 
\]
The constant of proportionality is $f(y_t|y_{1:t-1})$ and it does not depend on
$\theta$, but it does depend on the data. It is the one-step-ahead predictive
density of $y_t$. In this way, one can approximate the expectation
$\pi_t(\varphi) \coloneq \mathbb{E}_{\pi_t}[\varphi(\theta)]$ at time $t$, for a generic function $\varphi$ such that this expectation exists, with the weighted average 
\begin{equation}\label{eq:sis_estimate}
  \frac{1}{\sum_{n = 1}^{N_\theta} \omega_t^n} \times \sum_{l=1}^{N_\theta} \omega_t^l \varphi(\theta^l), 
\end{equation}

Furthermore, the average of the weights, $N_\theta^{-1}\sum_{l=1}^{N_\theta} \omega_t^l$, is an unbiased estimate of the marginal likelihood $f(y_{1:t})$, since 
$\omega_t^l = f(y_{1:t}|\theta^l)$ and the $\theta^l$ are drawn from the prior. The increments $f(y_t|y_{1:t-1}) = \pi_{t-1}(\varphi_t)$, with $\varphi_t(\theta) = f(y_t|y_{1:t-1},\theta)$, are the one-step-ahead predictive densities, and may in turn be estimated by applying \eqref{eq:sis_estimate} to $\varphi_t$ under the time $(t-1)$ weights. It is therefore possible to approximate recursively the marginal likelihood $\prod_{s=1}^t f(y_s|y_{1:s-1})$, and hence to perform sequential model choice. Restricted to the observations that follow a training sample, this product becomes the predictive likelihood routinely used to compare macroeconomic forecasting models \citep{GewekeAmisano2010}. One can also sample at time $t$ from the predictive distribution of $y_{t+1}$, by simulating $y_{t+1}^l \sim f(y_{t+1}|y_{1:t},\theta^l)$ for each $l\in[N_\theta]$, and reporting weighted averages over these $y_{t+1}^l$'s.

This naive approach may work for a few iterations, but it will eventually suffer from weight degeneracy: as data accumulates, the effective support of $\pi_t$, which is the region that contains most of its mass, shrinks and no longer matches that of $\pi_0$, so that only the few `particles' $\theta^l$ that happen to fall in this region retain non-negligible normalised weight, so that \eqref{eq:sis_estimate} is effectively an average over a handful of points and has a large variance. 


To counter degeneracy, one may, from time to time, trigger a resample-move step: one first resamples the particles (\textit{i.e.}, samples $N_{\theta}$ times with replacement, with probabilities proportional to the weights), which discards low-weight particles and duplicates high-weight ones, thereby refocusing the sample on the effective support of $\pi_t$. One then moves the particles through a few MCMC steps that leave $\pi_t$ invariant, which restores diversity among the duplicates. This move step is also called rejuvenation, and we use the two terms interchangeably.
 
More precisely, let $M$ and $P$ be two integers such that $N_{\theta} = M \times P$. In the waste-free variant \citep{wastefreeSMC} of IBIS, one resamples only a small number $M\ll N_{\theta}$ of the particles, and uses these $M$ particles as starting points for $M$ Markov chains of length $P$, using a Markov kernel that leaves invariant the current target distribution $\pi_t$. In this way, one gets a new, `fresh' sample (with weights set to one) that may be used in the subsequent reweighting steps, see Algorithm \ref{alg:ibis} for a summary. The remaining question is when to trigger such steps.

To determine when to perform a resample-move step, a standard strategy is to monitor the ESS (effective sample size) of the weights, which is defined as
$\left(\sum_{l=1}^{N_\theta} \omega_t^l\right)^2/\sum_{l=1}^{N_\theta} (\omega_t^l)^2$. This quantity lies in $[1, N_\theta]$.  The resample-move
steps are triggered at times $t$ where the ESS gets below $\gamma N_\theta$,
for some $\gamma\in(0, 1)$, \textit{e.g.},  $\gamma=1/2$. 

\begin{algorithm}
  \caption{Waste-free IBIS}\label{alg:ibis}
  \begin{algorithmic}
    \Require prior $\pi_0$, particle number $N_\theta = M\times P$, family of MCMC kernels $(\mathcal{K}_t)$ with $\mathcal{K}_t$ leaving $\pi_t$ invariant, ESS threshold $\gamma\in(0,1)$.
    \State $\theta^l \sim \pi_0$ and $\omega^l \gets 1$, for $l\in[N_\theta]$
      \hfill\Comment{Initialise from the prior}
    \For{$t=1,2,\dots$}
      \State report $\widehat{f}(y_t|y_{1:t-1}) \gets \left(\sum_{l=1}^{N_{\theta}} \omega^l\right)^{-1} \sum_{l=1}^{N_{\theta}} \omega^l\, u_t(\theta^l)$,\\
         \quad with $u_t(\theta) \coloneq f(y_t|y_{1:t-1}, \theta)$
        \hfill\Comment{Marginal likelihood; sequential model choice}
      \State $\omega^l \gets \omega^l \times u_t(\theta^l)$, for $l\in[N_\theta]$
        \hfill\Comment{Reweight}
      \State $\mathrm{ESS} \gets
        \left(\sum_{l=1}^{N_{\theta}} \omega^l\right)^2 / \sum_{l=1}^{N_{\theta}} (\omega^l)^2$
      \If{$\mathrm{ESS} < \gamma N_\theta$}
        \State draw $M$ indices $a_1,\dots,a_M$ with $\mathbb{P}(a_m = l) \propto \omega^l$, $l\in[N_{\theta}]$
          \hfill\Comment{Resample $M$ starting points}
        \For{$m=1,\dots,M$}
          \State run a Markov chain of length $P$ started at $\theta^{a_m}$ under $\mathcal{K}_t$; collect its $P$ states\\
            \hfill\Comment{Move}
        \EndFor
        \State relabel the $M\times P$ collected states as $\{\theta^l\}_{l\in[N_\theta]}$ and set $\omega^l \gets 1$ for $l\in [N_{\theta}]$\\
          \hfill\Comment{Fresh, equally weighted sample}
      \EndIf
    \EndFor
  \end{algorithmic}
  \medskip\footnotesize\emph{Note:} $u_t(\theta) \coloneq
  f(y_t|y_{1:t-1},\theta)$ denotes the incremental likelihood of $y_t$. The
  reported quantity $\widehat{f}(y_t|y_{1:t-1})$ estimates the one-step-ahead
  predictive density $f(y_t|y_{1:t-1})$, that is, the marginal likelihood
  increment. Its running product over $t$ yields the marginal likelihood
  $f(y_{1:t})$ used for sequential model choice. Each Markov chain of length
  $P$ comprises its starting point $\theta^{a_m}$ and $P-1$ successive draws
  from $\mathcal{K}_t$, so that the move step outputs $N_\theta = M \times P$
  states in total. At the end of each iteration $t$, the weighted system
  $\{\theta^l,\omega^l\}_{l\in[N_{\theta}]}$ approximates $\pi_t$. The
  algorithm thus delivers, on-line, the sequence of posterior approximations together with the estimates $\widehat{f}(y_t|y_{1:t-1})$.
\end{algorithm}

\subsection{\smcsq}\label{sub:smc2:generic}
IBIS requires, at each time $t$, the incremental likelihood $u_t(\theta)=f(y_t|y_{1:t-1},\theta)$ of the new datapoint. As noted above, this quantity is intractable for a state-space model, since it is an integral over the latent states. The \smcsq{} algorithm \citep{smc2} circumvents this by attaching to each parameter particle a \emph{local particle filter} that delivers an estimate of $u_t(\theta)$, and by using this estimate in place of the exact value in the IBIS reweighting step. In our model, the reparametrisation of Section \ref{sub:transformed_model} makes these local filters particularly cheap and stable: each decomposes into $K$ independent univariate bootstrap filters, one per transformed series $\wtl y_{1:t,k}$ (Section \ref{sub:smc2:recursive}), which keeps the variance of the resulting likelihood estimates under control with a moderate number of particles.

Concretely, one propagates $N_\theta$ parameter particles $\theta^l$ (as in IBIS), but each is now paired with a local particle filter of $N_x$ particles
$x_t^{l,n}$, $n\in[N_x]$, that tracks the filtering distribution of the states given $\theta=\theta^l$. Here, $N_x$ denotes the number of these `inner'
particles, as opposed to $N_\theta$, the number of `outer' (parameter) particles. At time $0$, one draws $\theta^l\sim\pi_0$ and initialises the $N_\theta$ local filters. Then, at each time $t$, one advances each local filter by one step, for instance one step of the bootstrap filter of Algorithm \ref{alg:bootstrap}: for each $l$ one samples states $x_t^{l,n}$ and computes weights $w_t^{l,n}$, in such a way that
\[
\widehat u_t(\theta^l) \coloneq \frac{1}{N_x}\sum_{n=1}^{N_x} w_t^{l,n}
\]
estimates $u_t(\theta^l) = f(y_t\mid y_{1:t-1},\theta^l)$. Each individual increment $\widehat u_t(\theta^l)$ is biased for $u_t(\theta^l)$ at finite $N_x$. The key property \citep{DelMoral1996unbiased} is that the running product $\prod_{s=1}^{t}\widehat u_s(\theta^l)$ is an unbiased estimate of the likelihood $f(y_{1:t}\mid\theta^l)$, for any $t$ and any fixed $N_x \geq 1$, \citep{DelMoral1996unbiased}. This is the property that the exactness of \smcsq{} requires. One then
updates the outer weights exactly as in IBIS, but with the estimate in place of
the exact value:
\[
  \omega_t^l \gets \omega_{t-1}^l \times \widehat u_t(\theta^l).
\]
Despite this substitution, \smcsq{} remains exact: the unbiasedness of the running product $\prod_{s\leq t}\widehat u_s$ ensures that the algorithm targets an extended distribution whose marginal in $\theta$ is the correct posterior $\pi_t(\theta) = \pi(\theta\mid y_{1:t})$ \citep{smc2}. In particular, the marginal-likelihood and predictive estimates of IBIS carry over unchanged, now driven by the $\widehat u_t(\theta^l)$, so that sequential model choice and forecasting are performed exactly as in IBIS.

As in IBIS, the outer weights eventually degenerate, and one monitors the ESS, triggering a resample-move step whenever it falls below $\gamma N_\theta$. The move step, however, requires more care than in IBIS. The invariance requirement is unchanged: the kernel $\mathcal{K}_t$ must leave the current target $\pi_t(\theta)$ invariant, but this target is the posterior distribution of $\theta$ with the latent states marginalised out. Any standard MCMC kernel targeting it would therefore require evaluating the intractable likelihood $f(y_{1:t}\mid\theta)$, the very quantity whose intractability motivated \smcsq{}. To rejuvenate the parameter particles, that is, to implement this move step, we therefore use a PMCMC kernel \citep{PMCMC}, which sidesteps the marginalisation by operating on the pair $(\theta, x_{1:t})$: each parameter particle is first augmented with a state trajectory $x_{1:t}^l$ extracted from its own local filter, and the pair is then moved by a Markov kernel that leaves invariant the extended target, whose $\theta$-marginal is $\pi_t(\theta)$, as required. In our setting this kernel is precisely the Particle Gibbs kernel of Section \ref{sec:ourMCMC}, with the $K$ volatility trajectories refreshed by the CSMC updates of Algorithm \ref{alg:CSMC}. After the move, the local filter of each rejuvenated pair is reinitialised, so that the reweighting steps can resume at time $t+1$.

Finally, we adopt the waste-free variant of this scheme \citep{wastefreeSMC}, exactly as we did for IBIS. This choice is not incidental: in our model $\theta=(A,\tilde\Pi,\Phi)$ has dimension $\bigO(K^2 p)$, so moving a particle meaningfully requires a non-trivial number of MCMC sweeps, and the waste-free scheme recycles every intermediate state of those sweeps rather than discarding all but the last. With $N_\theta = M\times P$, a resample-move step resamples only $M\ll N_\theta$ pairs $(\theta^l, x_{1:t}^l)$, and uses each as the starting point of a Markov chain of length $P$ under the PMCMC kernel. The $M\times P$ resulting states are collected, relabelled, and given unit weights. Algorithm~\ref{alg:wf-smc2} summarises the resulting waste-free \smcsq{} sampler.

\begin{algorithm}
  \caption{Waste-free \smcsq{}} \label{alg:wf-smc2}
  \begin{algorithmic}
    \Require prior $\pi_0$, outer particles $N_\theta = M\times P$, inner particles $N_x$, family of PMCMC kernels $(\mathcal{K}_t)$ with $\mathcal{K}_t$ leaving the extended target at time $t$ invariant, ESS threshold $\gamma\in(0,1)$.
    \State $\theta^l \sim \pi_0$ and $\omega^l \gets 1$, for $l\in[N_\theta]$
      \hfill\Comment{Initialise parameters from the prior}
    \State initialise a local particle filter of $N_x$ particles for each
      $\theta^l$
      \hfill\Comment{One inner filter per outer particle}
    \For{$t=1,2,\dots$}
      \For{$l=1,\dots,N_\theta$}
        \State advance the local filter of $\theta^l$ by one step: sample
          $x_t^{l,n}$ and weights $w_t^{l,n}$, $n\in[N_x]$\\
          \hfill\Comment{e.g.\ bootstrap filter, Alg.~\ref{alg:bootstrap}}
        \State $\widehat u_t(\theta^l) \gets
          N_x^{-1}\sum_{n=1}^{N_x} w_t^{l,n}$
          \hfill\Comment{Increment estimate. $\prod_{s\leq t}\widehat u_s(\theta^l)$ is unbiased for $f(y_{1:t}\mid\theta^l)$}
      \EndFor
      \State report $\widehat f(y_t|y_{1:t-1}) \gets
        \left(\sum_{l=1}^{N_{\theta}} \omega^l\right)^{-1} \sum_{l=1}^{N_{\theta}} \omega^l\, \widehat u_t(\theta^l)$
        \hfill\Comment{Marginal likelihood; sequential model choice}
      \State $\omega^l \gets \omega^l \times \widehat u_t(\theta^l)$,
        for $l\in[N_\theta]$
        \hfill\Comment{Reweight}
      \State $\mathrm{ESS} \gets
        \left(\sum_l \omega^l\right)^2 / \sum_l (\omega^l)^2$
      \If{$\mathrm{ESS} < \gamma N_\theta$}
        \State draw $M$ indices $a_1,\dots,a_M$ with
          $\mathbb{P}(a_m = l) \propto \omega^l$
          \hfill\Comment{Resample $M\ll N_\theta$ survivors}
        \For{$m=1,\dots,M$}
          \State extract a state trajectory $x_{1:t}^{a_m}$ from the local
            filter of $\theta^{a_m}$
            \hfill\Comment{Augment $\theta$ with a state path}
          \State run a Markov chain of length $P$ started at
            $(\theta^{a_m}, x_{1:t}^{a_m})$ under $\mathcal{K}_t$;
            collect its $P$ states
            \hfill\Comment{Move}
        \EndFor
        \State relabel the $M\times P$ collected pairs
          $(\theta^l, x_{1:t}^l)$ as $\{\theta^l\}_{l\in[N_\theta]}$ (with their
          local filters) and set $\omega^l \gets 1$
          \hfill\Comment{Fresh, equally weighted sample}
      \EndIf
    \EndFor
  \end{algorithmic}
  \medskip
  \footnotesize\emph{Note:} Notation as in Algorithm \ref{alg:ibis}. A trajectory $x_{1:t}^l$ is extracted from a local filter by drawing a terminal index $b_t \sim \mathrm{Cat}(W_t^{l,1:N_x})$ and tracing back the ancestors, $b_s = a_{s+1}^{b_{s+1}}$. Each length-$P$ chain operates on pairs $(\theta, x_{1:t})$, its starting pair included; after relabelling, each pair's local filter is reinitialised (see Algorithm~\ref{alg:smcsq}). At the end of iteration $t$, $\{(\theta^l,\omega^l)\}$ approximates $\pi(\theta\mid y_{1:t})$, and the running product of the reported $\widehat f(y_t\mid y_{1:t-1})$, the one-step-ahead predictive densities, estimates the marginal likelihood $f(y_{1:t})$.
\end{algorithm}

The basic waste-free \smcsq{} sampler above requires several modifications to adapt it to our context. We detail them in the three subsections below:
initialisation from a training sample rather than from the prior (Section~\ref{sub:smc2:init}); recursive updates that assimilate the data one scalar component of the transformed observation vector at a time (Section~\ref{sub:smc2:recursive}); and rejuvenation under partial observations (Section~\ref{sub:smc2:rejuv}). Algorithm~\ref{alg:smcsq}, given at the end of the section, summarises the resulting sampler as used in this paper.

\subsubsection{Initialisation from a training sample}\label{sub:smc2:init}

The textbook exposition above starts the sampler at $t=0$ by drawing $\theta^l\sim\pi_0$ from the prior. In our application we deliberately depart from this: we condition on a training (or pre-) sample $y_{1:T_i}$ and initialise the population at $t=T_i$ from the training-sample posterior $\pi(\theta\mid y_{1:T_i})$. Concretely, we draw the initial population of $N_\theta$ particles by running the Particle Gibbs sampler of Section \ref{sec:ourMCMC} on $y_{1:T_i}$: we launch $M$ independent chains, discard the first $N_{\text{burn}}$ sweeps of each, and retain $P$ consecutive post-burn-in sweeps per chain. This reproduces exactly the waste-free block geometry $N_\theta = M\cdot P$ used throughout Section \ref{sec:SMC2}: the initial population has the same dependence structure ($M$ chains of length $P$) as the population produced by any later resample-move step. Each retained sweep supplies both a parameter draw $\theta^l$ and an associated volatility path drawn, via the CSMC updates, from the smoothing distribution $\pi(\lambda_{1:T_i}\mid\theta^l, y_{1:T_i})$. The terminal value $\lambda_{T_i}^l$ of each retained trajectory is a draw from the filtering distribution $\pi(\lambda_{T_i}\mid \theta^l, y_{1:T_i})$, so each local filter starts the on-line phase from a volatility level informed by the full training sample, rather than from the prior on $\lambda_0$, which reflects no data. Concretely, the local filter of particle $l$ is initialised by propagating its $N_x$ particles from the single retained value $\lambda_{T_i}^l$ through the volatility transition \eqref{eq:same_motion_model}. This is the same reinitialisation used after every resample-move step, and it is valid because the retained pair $(\theta^l, \lambda_{1:T_i}^l)$ is itself part of the extended target. 
In the real-data experiment of Section \ref{sec:num} we use $T_i=236$, $N_{\text{burn}}=500$, $N_\theta=7500$ and $P=500$ (hence $M=15$), and start the on-line recursion at $t=T_i+1$.

Two considerations motivate this choice. First, it is the fully Bayesian counterpart of two standard practices: calibrating priors on an initial training sample, as in \citet{Primiceri2005}, who sets the priors on initial states and hyperparameters from the first ten years of data, and evaluating models by their predictive performance only after an initial estimation window, as is routine in out-of-sample forecast evaluation. Whereas the former plugs training-sample point estimates into the prior, we carry the exact training-sample posterior forward: $\pi(\theta\mid y_{1:T_i})$, which is proper by construction, plays the role of the prior for the remaining data. This is an application of Bayes' rule at $t=T_i$, with no approximation. The same device, using a training-sample posterior as a proper prior for marginal-likelihood-based model comparison, is employed in the moment-condition literature by \citet{ChibShinSimoni2022}.

Second, and more fundamentally, model comparison is based on the predictive likelihood $\prod_{t>T_i} f(y_t\mid y_{1:t-1})$, evaluated on a training sample common to all models. Ratios of these quantities across models are Bayes factors conditional on $y_{1:T_i}$, in the spirit of the predictive-likelihood comparisons of \citet{GewekeAmisano2010}. Equivalently, this quantity is the marginal likelihood of $y_{T_i+1:t}$ conditional on the training sample, $f(y_{T_i+1:t}\mid y_{1:T_i}) = f(y_{1:t})/f(y_{1:T_i})$: the algorithm computes the full marginal likelihood $f(y_{1:t})$ when initialised at $\pi_0$, and its training-sample-conditional counterpart when initialised at $\pi(\theta\mid y_{1:T_i})$, as we do throughout the application. This construction has two benefits. It yields less fragile model comparison: with $\bigO(K^2 p)$ coefficients, the marginal likelihood computed from $\pi_0$ would be dominated by the arbitrary dispersion of its vague components (such as the $\mathcal{N}(0, 10^6)$ prior on the free elements of $A$) through the Lindley-Bartlett effect (\cite{Lindley1957, Bartlett1957}), whereas conditioning on the training sample replaces this dispersion with a data-determined one, common to all models compared. It also improves numerical robustness: starting at $t=T_i$ bypasses the early, diffuse regime in which the importance weights collapse and the ESS degenerates almost immediately (the same failure mode that motivated the tight prior on $\Phi$ in Section \ref{sec:model}). The two devices work in tandem. Bypassing this regime matters more here than in low-dimensional applications: with $\theta=(A,\wtl\Pi,\Phi)$ of dimension $\bigO(K^2 p)$, each rejuvenation is a full Particle Gibbs pass over the accumulated data, so the near-continuous rejuvenation that a prior-initialised run would trigger in its early phase
would dominate the total cost.

\subsubsection{Recursive updates one observation at a time}\label{sub:smc2:recursive}
The reparametrisation $\wtl{y}=Ay$ of Section \ref{sub:transformed_model} lets us assimilate each observation vector $y_t$ one scalar component at a time: within date $t$, the $K$ rows of the transformed system \eqref{model:eq:reparametrized} are processed sequentially, for $k\in[K]$, the $k$-th row being the univariate SV equation that relates $\wtl y_{t,k}$ to its own volatility $\lambda_{t,k}$. As explained in Section \ref{sec:relevance:our:model}, the $K$ volatility trajectories are conditionally independent given $\theta$, so the generic local particle filter of Section \ref{sub:smc2:generic} is in fact a collection of $K$ independent univariate SV bootstrap filters (Algorithm \ref{alg:bootstrap}), the $k$-th run on the scalar series $\wtl{y}_{1:t,k}=(Ay)_{1:t,k}$. Moreover, since $A$ is unit lower-triangular, the map $y_t\mapsto\wtl y_t=Ay_t$ has unit Jacobian and $\wtl y_{t,k}=y_{t,k}+\sum_{j<k}A_{k,j}y_{t,j}$ depends only on $y_{t,1:k}$. The date-$t$ density therefore factorises into $K$ scalar conditionals with no Jacobian correction, the $k$-th of which conditions on the partial current observation $y_{t,1:k-1}$ (equivalently $\wtl y_{t,1:k-1}$) through the mean shift $\sum_{j<k}A_{k,j}y_{t,j}$. This per-equation assimilation is the sequential counterpart of the equation-by-equation treatment of large triangularised VAR-SV systems in \citet{CARRIERO2019137} and \citet{CarrieroChanClarkMarcellino2022}, transposed from the Gibbs sampler to the likelihood-estimation layer of
\smcsq{}. Unlike there, our per-equation terms are fully decoupled given $\theta$ (Section \ref{sec:relevance:our:model}).

Concretely, within date $t$ we loop over $k\in[K]$, running the $k$-th univariate filter one step and updating the outer weight at each sub-step $(t,k)$,
\[
  \omega^l \;\gets\; \omega^l \times \widehat f_{t,k}^{\,l}, \qquad
  \widehat f_{t,k}^{\,l} \coloneq \frac{1}{N_x}\sum_{n=1}^{N_x} w_{t,k}^{l,n},
\]
which estimates
$f\!\left(\wtl y_{t,k}\mid\theta^l, y_{1:t-1},y_{t,1:k-1}\right)$.


As in Section \ref{sub:smc2:generic}, each individual increment $\widehat f_{t,k}^{\,l}$ is, in general, biased for its target at finite $N_x$ (the first observation assimilated by each filter being the lone exception). The property that survives, and all that the exactness of \smcsq{} requires, is unbiasedness of the running product: for each $k$, the $k$-th filter satisfies $\EE\big[\prod_{s\le t}\widehat f_{s,k}^{\,l}\big] =\prod_{s\leq t} f(\wtl y_{s,k}\mid y_{1:s-1},\theta^l)$ \citep{DelMoral1996unbiased}, and, since the $K$ filters are driven by independent random numbers given $\theta^l$, the running product over all assimilated sub-steps,
\[
  \prod_{j=1}^{K}\prod_{s=1}^{t-1}\widehat f_{s,j}^{\,l}\;\times\;\prod_{j=1}^{k}\widehat f_{t,j}^{\,l},
\]
is unbiased for the partial-data likelihood $f(y_{1:t-1},\,y_{t,1:k}\mid\theta^l)$, for any $(t,k)$ and any fixed $N_x \geq 1$. Because unbiasedness holds at every sub-step, and not merely at date boundaries, the extended-target argument of Section \ref{sub:smc2:generic} applies verbatim after each equation is assimilated: the weighted parameter particles target $\pi(\theta\mid y_{1:t-1}, y_{t,1:k})$. A single time step thus decomposes into $K$ sub-steps, with the ESS monitored at each $(t,k)$. This is what allows a resample-move step to be triggered, and handled exactly, when the current date is only partially observed (Section \ref{sub:smc2:rejuv}). In the training-sample-initialised sampler of Section \ref{sub:smc2:init}, the products run over the sub-steps after $T_i$ and the same statement holds conditionally on the retained pair $(\theta^l,\lambda_{T_i}^l)$, which is itself part of the extended target. Their outer aggregate is the estimate $\widehat f(y_{T_i+1:t}\mid y_{1:T_i})$ reported by Algorithm \ref{alg:smcsq}.

\subsubsection{Rejuvenation under partial observations}\label{sub:smc2:rejuv}
In the \smcsq{} recursion of Section \ref{sub:smc2:recursive}, each new observation vector $y_t$ is processed one equation at a time. The ESS may therefore drop below its threshold when only the first $k<K$ equations of date $t$ have been processed. At that moment a resample-move step starts: $y_{t,1:k}$ has been processed while $y_{t,k+1:K}$ has not. Equivalently, the equations have different sample sizes: the first $k$ equations have $t$ observations, the remaining $K-k$ have $t-1$. We stress that this situation arises even when the dataset is complete. The date is partially observed from the algorithm's point of view, simply because the move occurs before its last $K-k$ equations have been processed. The same situation also arises when the final date of the sample is genuinely incomplete, because the last $K-k$ series in the model's ordering have not yet been released. All earlier dates must be fully observed.

In both cases the rejuvenation kernel must target $\pi(\theta\mid y_{1:t-1}, y_{t,1:k})$. More precisely, as in Section \ref{sub:smc2:generic}, the kernel operates on an extended target whose $\theta$-marginal is this posterior. After the move, filter $j$ restarts at date $t$ if $j \leq k$ and at date $t-1$ otherwise. Every parameter block copes with the missing tail in the same way: the data sums of the unobserved equations $j>k$ stop at $t-1$, while those of the observed equations $j\leq k$ run to $t$, so that no missing value is imputed. The standard treatment of an incomplete final date, the ``ragged edge'' of real-time data \citep{Wallis1986}, is instead data augmentation. The missing observations are simulated within the sampler, as in nowcasting and mixed-frequency applications \citep{SchorfheideSong2015,CarrieroClarkMarcellino2015}. Our truncation removes this step as the unreleased series are ordered last by construction. For the transformed coefficients, the volatilities and $\Phi$ the truncation is immediate. For $A$ a short additional argument is needed, which we sketch below and give in full in Appendix \ref{app:A_partial}.

\paragraph{Parameters that need no imputation ($\wtl{\Pi}$, $\lambda$, $\Phi$).}
Recall that our sampler updates the transformed coefficients $\wtl{\Pi}=A\Pi$ rather than $\Pi$ itself. In these three blocks, $A$ is held fixed. It is
updated only in its own block, described next. Given $A$, the matrices $\wtl{\Pi}=A\Pi$ and $\Pi=A^{-1}\wtl{\Pi}$ are in one-to-one correspondence, so a valid draw of one is a valid draw of the other, and we may work in whichever coordinates are convenient. We use $\wtl{\Pi}$ because in this coordinate the
likelihood factorises across equations. The cross-equations coupling resides entirely in the prior, which involves the parameters but no data (Section \ref{sec:ourMCMC}). This separation is what removes any need for imputation. The no-imputation property established next therefore holds for the $\Pi$ block as well, at no extra cost, because a draw of $\wtl{\Pi}$ is a draw of $\Pi$. 

In particular, given $A$, the likelihood terms of equation $j$ involve only its own coefficients and the observations $y_{t',1:j}$, $t'\leq t$ (Section \ref{sec:ourMCMC}). A missing tail $y_{t,k+1:K}$ therefore enters only the unobserved equations $j>k$. For those it is enough to stop the data sums at $t-1$. The observed equations $j\leq k$ are not affected. This is the no-imputation property of Remark \ref{rem:no_augmentation}, stated there with the \smcsq{} sampler in mind. The full conditional of $\wtl{\Pi}_{(j)}$ from Section \ref{sec:ourMCMC} is thus reused unchanged, except that its data sums run to $t$ for the observed equations $j\leq k$ and to $t-1$ for the unobserved equations $j>k$. The CSMC update of $\lambda_{\cdot,j}$ (Algorithm \ref{alg:CSMC}) and the inverse-Gamma update of $\Phi_{j,j}$ use the same per-equation sample sizes. No missing value is imputed in these blocks. At the end of the sweep, $\Pi = A^{-1}\wtl{\Pi}$ is recovered without ever filling in the missing part of $y_t$.

\paragraph{The matrix $A$.}
The matrix $A$ may seem to be an exception, since it is itself the map $y_t\mapsto\wtl{y}_t = Ay_t$ that produces the transformed data, so its update apparently involves the full vector $y_t$. In fact no imputation is needed here either. The same truncation applies, row by row, because date $t$ contributes no term to the update of the rows $A_{(j)}$ with $j>k$. To see this, consider the factors of the period-$t$ likelihood that involve the missing observations $y_{t,k+1:K}$, and change variables from these observations to the associated structural residuals $r_{t,k+1:K}$. Conditionally on the observed $y_{t,1:k}$, this change of variables is triangular with unit diagonal, hence has unit Jacobian, so each factor becomes a Gaussian density in its own residual and integrates to one, for any value of the volatilities. Integrating the volatilities $\lambda_{t,k+1:K}$ over their transition densities also gives one. The period-$t$ contribution of the missing block therefore equals one, whatever the value of $A$, and drops from the update. The data sums for row $A_{(j)}$ thus run to $t$ when $j \leq k$ and to $t-1$ when $j>k$, with no imputation. Updating a row $j \leq k$ additionally requires the contemporaneous reduced-form residuals $v_{t,1:j-1}$. These are observed too, again because $A$ is lower triangular. Appendix~\ref{app:A_partial} gives the full argument.

\subsubsection{The complete sampler}\label{sub:smc2:algo}

Combining the three modifications above yields the sampler we use in this paper, summarised in Algorithm \ref{alg:smcsq}. As in the generic scheme, the particles are rejuvenated by a move step on an extended space. Each particle $\theta^l$, with $\theta=(A,\wtl{\Pi},\Phi)$, is augmented by a state trajectory $x^l=\log\lambda^l$ drawn from its own local filters, with the per-equation lengths of Section \ref{sub:smc2:rejuv}. The augmented pair is then rejuvenated by the kernel of Section \ref{sec:ourMCMC}, with $\lambda$ updated by the CSMC step of Algorithm \ref{alg:CSMC}.


\begin{algorithm}
  \caption{Waste-free SMC$^2$ (initialisation from a Particle Gibbs presample; rejuvenation via the kernel of Section \ref{sec:ourMCMC})}
  \label{alg:smcsq}
  \begin{algorithmic}
    \Require training-sample cutoff $T_i$, data $y_{1:T}$, $K$ equations, $N_\theta = M\times P$ outer / $N_x$ inner particles, ESS threshold $\gamma$.
    \State initialise $\{\theta^l, x_{1:T_i}^{l,1:N_x}\}_{l\in[N_\theta]}$ and $\omega^l\gets1$ by running $M$ Particle Gibbs chains (\S\ref{sec:ourMCMC}) on $y_{1:T_i}$ and keeping $P$ consecutive post-burn-in sweeps each
      \hfill\Comment{Training-sample init, $N_\theta=M\times P$; see \S\ref{sub:smc2:init}}
    \For{$t=T_i+1, T_i+2, \dots$}
      \For{$k=1,\dots,K$}
        \hfill\Comment{Assimilate one row of \eqref{model:eq:reparametrized} at a time}
        \For{$l=1,\dots,N_\theta$}
          \State propagate $x_{t,k}^{l,n}\sim \pi^{\theta^l}(\cdot\mid x_{t-1,k}^{l,a})$ 
          \State and weight $w_{t,k}^{l,n}=f(\wtl y_{t,k}\mid\theta^l,x_{t,k}^{l,n})$, $n\in[N_x]$
          \hfill\Comment{only filter $k$ advances, Alg.~\ref{alg:bootstrap}}
          \State normalise, resample $a_{t,k}^{l,n}$, and set $\widehat f_{t,k}^l \gets N_x^{-1}\sum_n w_{t,k}^{l,n}$
          \State $\omega^l \gets \omega^l \times \widehat f_{t,k}^l$
          \hfill\Comment{$\widehat f_{t,k}^l$ estimates $f(\wtl y_{t,k}\mid\theta^l,y_{1:t-1},y_{t,1:k-1})$}
        \EndFor
        \State $\mathrm{ESS}\gets(\sum_l\omega^l)^2/\sum_l(\omega^l)^2$
        \If{$\mathrm{ESS} < \gamma N_\theta$}
        
        \State resample $M$ survivors $\propto \omega^l$; for each, draw trajectories 
        \State $x_{1:t,j}^l$ for $j\leq k$ and $x_{1:t-1,j}^l$ for $j>k$ from its local filters
        \hfill\Comment{Augment each survivor with state paths; \S\ref{sub:smc2:rejuv}}
        
        \State from each survivor run a length-$P$ chain of the kernel of 
        \State Section \ref{sec:ourMCMC} (update $A,\wtl{\Pi},\Phi,\lambda_{1:T}$; $\lambda_{1:T}$ via Alg.\ref{alg:CSMC})
        \State given $(y_{1:t-1},y_{t,1:k})$
        \hfill\Comment{Waste-free: keep all $M\times P$ states}

        \State relabel the $M\times P$ pairs, reinitialise each pair's $K$ local filters 
        \State from the terminal values of its trajectories, and set $\omega^l\gets 1$
        \hfill\Comment{Reweighting resumes at sub-step $(t,k+1)$}
        \EndIf
      \EndFor
    \EndFor
  \end{algorithmic}
  \medskip
  \footnotesize\emph{Note:} Notation and conventions as in Algorithms \ref{alg:ibis} and \ref{alg:wf-smc2}. Here, $\theta = (A, \wtl{\Pi}, \Phi)$ and $x^l = \log\lambda^l$. Each outer particle carries $K$ independent univariate filters, the $k$-th run on $\wtl y_{1:t,k} = (Ay)_{1:t,k}$. The increment estimates $\widehat f^{l}_{t,k}$ are individually biased in general. Their running product over all assimilated sub-steps is unbiased for the corresponding partial-data likelihood (Section \ref{sub:smc2:recursive}). A resample-move step triggered at equation $k$ targets $\pi(\theta, x \mid y_{1:t-1},y_{t,1:k})$ via the truncated updates of Section \ref{sub:smc2:rejuv}, with $\lambda$ updated by $K$ CSMC steps (Algorithm \ref{alg:CSMC}). For $k=K$ these coincide with the complete-data updates. The reported increments are estimates of the one-step-ahead predictive densities and are used for on-line forecasting. Their running product, $\widehat f(y_{T_i+1:t}\mid y_{1:T_i})$, estimates the predictive likelihood of the post-training data and is used for model choice.
\end{algorithm}

\subsection{Memory cost}\label{sub:smc2:memory}

A basic implementation of Algorithm~\ref{alg:wf-smc2} may be memory intensive, as it requires keeping in memory $N_\theta \times N_x$ state trajectories
$x_{1:t}^{l,n}$, amounting to $N_\theta \times N_x \times t \times K$ $\lambda$-variables at time $t$ in our model (since $\lambda_t$ is of dimension $K$). 

The memory footprint of this algorithm can be significantly reduced by using
the genealogy trick of \citet{pathstorage}: since, for each $\theta^l$, at most
one state trajectory is ever extracted from the set of particles
$x_{1:t}^{1:N_x}$, one can dispose of any particle $x_s^{n, l}$, $s<t$, that
has no descendant at time $t$. The aforementioned paper shows that storing only
particles with a descendant has average cost $\bigO(N_x \log N_x)$ instead of
$\bigO(N_x t)$, because of a coalescence effect. Hence, as soon as $t$ is large
enough, it is beneficial to implement this strategy. Then the overall memory
complexity of Algorithm~\ref{alg:wf-smc2} is $\bigO\left(K N_\theta N_x \log
N_x\right)$. 

This bound covers the reweighting phase, during which at most one trajectory is ever extracted from each local filter, by ancestral tracing. It does not, by itself, cover the move step, and the two devices we use interact here. The backward pass of Algorithm \ref{alg:CSMC} draws $b_s$ with probabilities proportional to $W_s^{n}\, p_\theta(x_{s+1}^{b_{s+1}}\mid x_s^{n})$ over \emph{all} $n\in[N_x]$, so it may revive a particle with no descendant at time $t$: the surviving genealogy does not suffice, and a CSMC step must hold the whole cloud $x_{1:t}^{1:N_x}$, at cost $\bigO(K N_x t)$ per chain. The waste-free geometry is what bounds the number of clouds held at once by $M \ll N_\theta$ rather than by $N_\theta$, keeping the move within the bound above as long as $M t = \bigO(N_\theta \log N_x)$, that is $t =\bigO(P \log N_x)$, a condition amply satisfied by the settings of Section \ref{sec:num}. The same length-$P$ chains also repair the coalescence that the genealogy trick exploits: the trajectory extracted by ancestral tracing is shared across the $N_x$ stored paths at early dates, and it is the succession of backward-sampling CSMC steps that restores diversity in those coordinates. The low memory footprint of Algorithm \ref{alg:smcsq} is thus not attributable to either device alone, but to their combination.

\section{Numerical experiments}\label{sec:num}

%
%
%
%
%
%
\subsection{Experiment for MCMC}\label{sec:experiment mcmc}
In this section, we compare our particle Gibbs sampler with the algorithm of \citet{CARRIERO2019137}, as amended in \citet{CarrieroChanClarkMarcellino2022}, which we adapt to our model specification and use as the baseline. The two samplers differ in two blocks. For the VAR coefficients, our sampler draws the reparametrised $\wtl{\Pi} = A\Pi$ (Section \ref{sub:transformed_model}), whereas the baseline draws $\Pi$ with the corrected triangular algorithm (CTA). 
For the log-volatilities, our sampler uses the CSMC update with backward sampling (Section \ref{sec:pmcmc}), whereas the baseline uses the auxiliary-mixture sampler of \citet{KimShephardChib1998}, as prescribed in \citet{CARRIERO2019137}: the seven-component mixture approximation, with the mixture indicators redrawn within each sweep from their full conditional and all remaining blocks drawn marginally of them. 
%
%
All remaining blocks are identical across the two samplers. Since the two samplers differ in both the $\Pi$ and the volatility updates, the comparisons below contrast the samplers as a whole: Table \ref{tab:pi_mixing} contrasts the draws of the VAR coefficients and Table \ref{tab:log_lambda_mixing} the draws of the volatility states, and each contrast reflects the combined effect of the two modifications, as the mixing of each block also depends on the conditioning draws produced by the other.


We focus on two diagnostics: sample autocorrelation functions of the draws, which measure mixing per iteration, and the effective sample size per second (ESS/sec), which also accounts for running time. 


The data are $K=15$ monthly series from the FRED-MD database \citep{McCrackenNg2016}, from 1970:M1 to 2021:M3 ($T = 615$ raw observations). Details for variable choice and transformation are presented in Table \ref{tab:pg_data_set}. We set the lag length to $p=9$. After transforming the data and conditioning on the $p$ initial lags, the estimation sample has $T=605$ observations.


The CSMC update uses $N=100$ particles with backward sampling. Each sampler runs for 20,000 iterations and all results use the 15,000 draws left after a burn-in of 5,000 iterations, with no thinning.

\begin{table}[htbp]
\centering
\caption{Macroeconomic Variables and Transformations}
\label{tab:pg_data_set}
\begin{tabular}{@{}llll@{}}
\toprule
\textbf{FRED Mnemonic} & \textbf{Description} & \textbf{Transformation} \\ \midrule
RPI & Real Personal Income  & $\Delta \ln y_t \times 1200$ \\
INDPRO & Industrial Production Index & $\Delta \ln y_t \times 1200$ \\
CUMFNS & Capacity Utilization: Manufacturing & None \\
UNRATE & Civilian Unemployment Rate  & None \\
PAYEMS & All Employees, Nonfarm  & $\Delta \ln y_t \times 1200$ \\
CES0600000007 & Avg. Weekly Hours: Goods-Prod.  & None \\
CES0600000008 & Avg. Hourly Earnings: Goods-Prod.  & $\Delta \ln y_t \times 1200$ \\
HOUST & Housing Starts: Total New Privately Owned  & $\ln y_t$ \\
DPCERA3M086SBEA & Real Personal Consumption Expenditures  & $\Delta \ln y_t \times 1200$ \\
EXUSUK & U.S. / U.K. Foreign Exchange Rate  & $\Delta \ln y_t \times 1200$ \\
GS5 & 5-Year Treasury Constant Maturity Rate  & None \\
GS10 & 10-Year Treasury Constant Maturity Rate  & None \\
BAAFFM & Moody's Baa Corp. Bond Minus Fed Funds  & None \\
WPSFD49207 & Producer Price Index: Finished Goods  & $\Delta \ln y_t \times 1200$ \\
PCEPI & Personal Consumption Expenditures Index & $\Delta \ln y_t \times 1200$ \\ \bottomrule
\end{tabular}
\end{table}

\subsubsection{MCMC Mixing Performance}

\begin{figure}[htbp]
    \centering
    \includegraphics[width=0.8\linewidth]{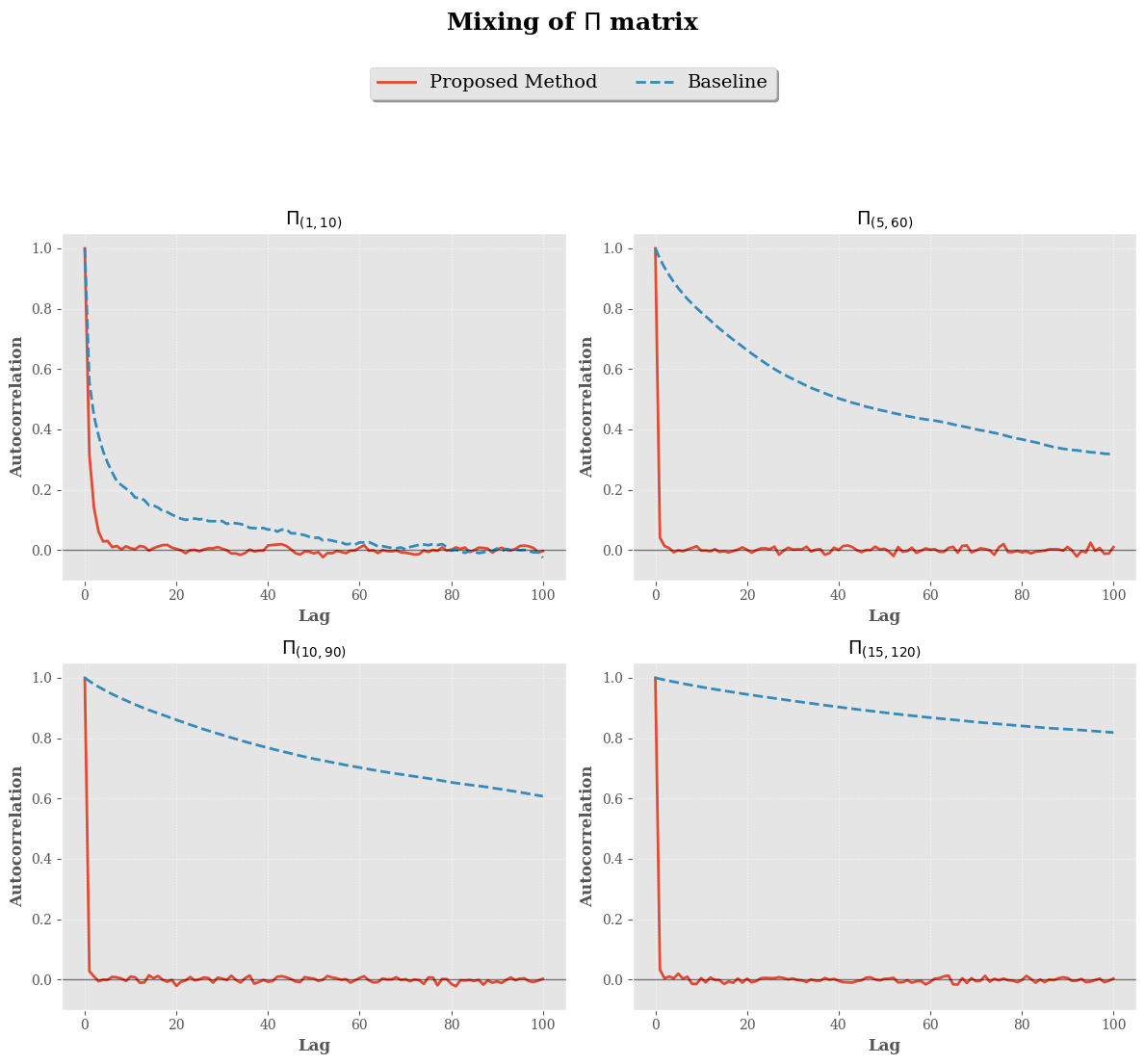}
    \caption{Sample autocorrelation functions of the MCMC draws for four elements of $\Pi$, under our sampler (reparametrised $\wtl{\Pi}=A\Pi$) and the baseline (CTA). FRED-MD data, $K=15$, $p=9$. 15,000 retained draws.}
    \label{fig:pi_mixing}
\end{figure}

As illustrated in Figure \ref{fig:pi_mixing}, the baseline method (the CTA) for updating $\Pi$ mixes slowly and so it exhibits significant stickiness, particularly for coefficients corresponding to later variables in the system (\textit{i.e.}, elements with larger row indices). In contrast, by reparameterizing $A\Pi$, we achieve consistently low autocorrelation across the entire coefficient matrix. Figure \ref{fig:pi_mixing} shows four elements of $\Pi$: $\Pi_{1,10}$ (the coefficient of the real income equation, RPI, on lag $1$ of real consumption), $\Pi_{5,60}$ (payroll employment, PAYEMS, on lag $4$ of the producer price index), $\Pi_{10,90}$ (the exchange rate, EXUSUK, on lag $6$ of the producer price index), and $\Pi_{15,120}$ (PCE inflation, PCEPI, on
lag $8$ of the producer price index). The four equations cover the main blocks of a standard medium-scale macroeconomic VAR (real activity, the labor market, financial variables, and prices) and span the recursive ordering from the first equation to the last. The columns are chosen at increasing lags, along which the mixing of the CTA deteriorates. Because equation index and lag increase together across the four panels, they display the combined effect. Figure \ref{fig:ACF:Pi} in Appendix \ref{App:figures} separates the two by reporting all $K=15$ equations at five fixed columns.

\begin{figure}[htbp]
    \centering
    \includegraphics[width=0.8\linewidth]{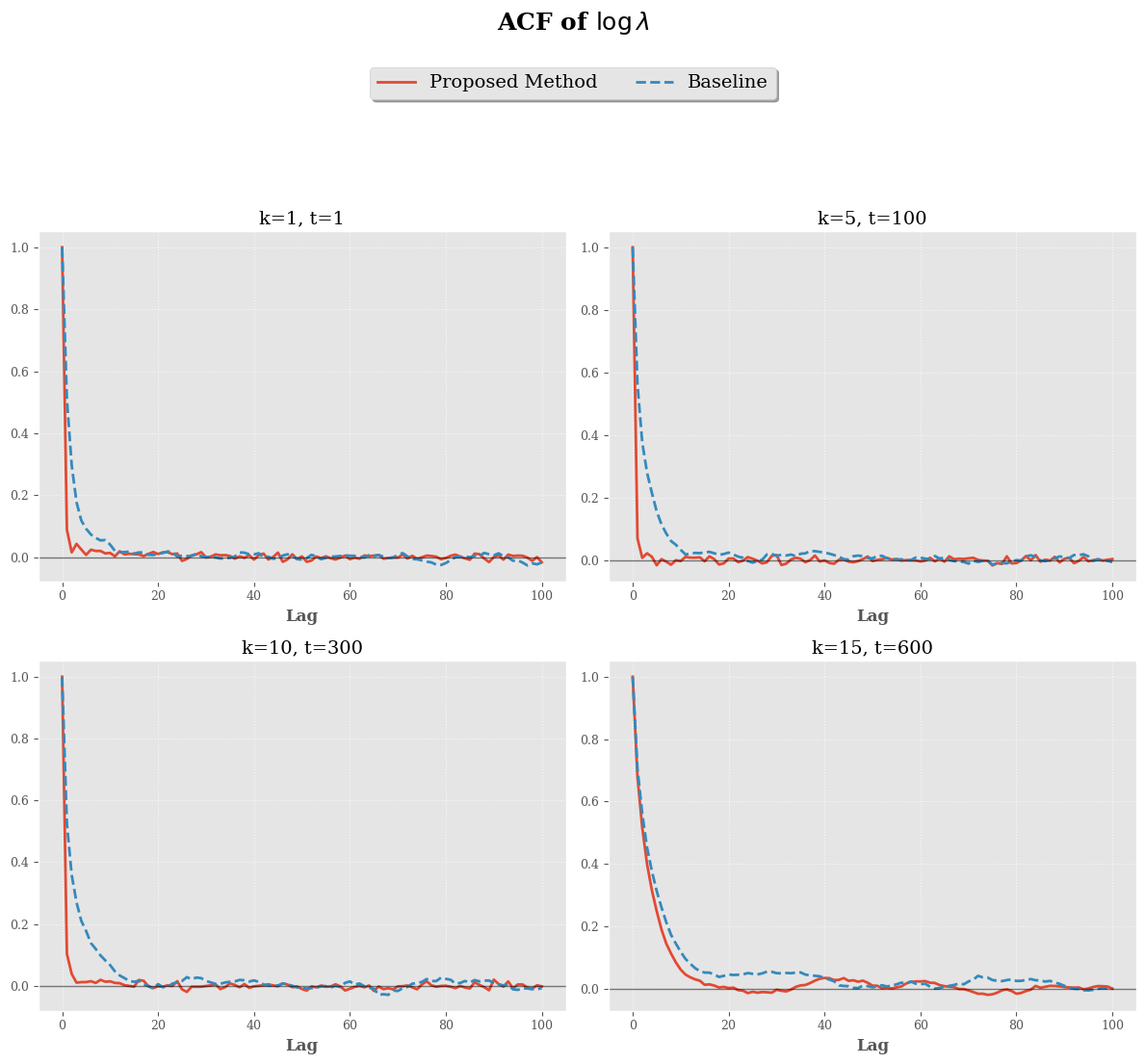}
    \caption{Sample autocorrelation functions of the MCMC draws for four latent log-volatility states $\log\lambda_{t,k}$, under our sampler (CSMC) and the baseline (KSC). Panels are labelled by series index $k$ and period $t$. Same data and settings as Figure \ref{fig:pi_mixing}.}
    \label{fig:loglambda_mixing}
\end{figure}

%
Figure \ref{fig:loglambda_mixing} reports the same diagnostics for the log-volatility paths. The CSMC update mixes better than the KSC mixture sampler at almost all coordinates. The exceptions are two states at the start of the sample, capacity utilization (CUMFNS) and PCE inflation (PCEPI) at $t=1$, visible in Figure \ref{fig:ACF:log:lambda}, which drive the low minimum ESS/sec in Table \ref{tab:log_lambda_mixing}. At all other coordinates CSMC matches or dominates KSC. Recall also that CSMC targets the exact posterior, whereas KSC targets the posterior of the auxiliary-mixture approximation.


Figure \ref{fig:loglambda_mixing} shows the autocorrelation functions for four states $\log\lambda_{t,k}$, with indices $(k,t)$ equal to $(1,1)$,
$(5,100)$, $(10,300)$, and $(15,600)$. The series are the same four equations as in Figure \ref{fig:pi_mixing} (RPI, PAYEMS, EXUSUK, PCEPI), covering the main blocks of the VAR, and the periods span the sample, from its start ($t=1$), through the late 1970s ($t=100$) and the calm mid-1990s ($t=300$), to the pandemic ($t=600$). Autocorrelation functions for a broader selection of states are reported in Figure \ref{fig:ACF:log:lambda} in Appendix \ref{App:figures} and display the same pattern.

Tables \ref{tab:pi_mixing} and \ref{tab:log_lambda_mixing} summarise ESS per second for VAR coefficients and for $\log\lambda_{t,k}$, respectively. Under our sampler, the mean ESS/sec is higher by a factor of about $14$ for the VAR coefficients and of about $3.4$ for the volatility states, relative to the baseline.


\begin{table}[htbp]
\centering
\caption{MCMC ESS per Second for VAR Coefficients (elements of $\Pi$), summarised over all $d_\Pi = K(Kp+1) = 2,040$ coefficients. LQ and UQ denote the 25th and 75th percentiles. Computed from 15,000 retained draws. CTA denotes the corrected triangular algorithm of \citet{CarrieroChanClarkMarcellino2022}.}\label{tab:pi_mixing}
\begin{tabular}{@{}lcccc@{}}
\toprule
Method & Mean  & Median & [LQ, UQ] & Range [Min, Max] \\
\midrule
$\wtl\Pi$ (ours) & 1.478323 & 1.521725 & [1.347025, 1.648324] & [0.478044, 1.848928] \\
$\Pi$ (Baseline with CTA) & 0.105182 & 0.080703 & [0.031748, 0.149994] & [0.000827, 0.674140] \\
\bottomrule
\end{tabular}
\end{table}

\begin{table}[htbp]
\centering
\caption{MCMC ESS per Second for Latent Volatility States ($\log\lambda_{t,k}$), summarised over all $K \times T$ coordinates. LQ and UQ denote the 25th and
75th percentiles. Computed from 15,000 retained draws}
\label{tab:log_lambda_mixing}
\begin{tabular}{@{}lcccc@{}}
\toprule
Method & Mean & Median & [LQ, UQ] & Range [Min, Max] \\
\midrule
CSMC (ours) & 0.997762 & 1.035259 & [0.808210, 1.228598] & [0.000895, 1.744017] \\
KSC (Baseline) & 0.293910 & 0.290412 & [0.208000, 0.380855] & [0.024367, 0.723851] \\
\bottomrule
\end{tabular}
\end{table}

\subsection{Experiment for $\text{SMC}^2$}\label{sec:experiment smcsquare}

In this section, we present the experimental results for the $\text{SMC}^2$ algorithm. The number of parameters in the VAR model grows $\mathcal{O}(K^2 p)$ with the number of time series $K$, but SMC algorithms in high dimensions suffer from particle degeneracy. We therefore restrict our empirical evaluation to smaller values of $K$. We evaluate the performance of $\text{SMC}^2$ along two dimensions. The first is rejuvenation efficiency, measured by how often the particle system needs to be rejuvenated, which we read from the ESS trace. The second is model selection, based on the estimated marginal likelihood. SMC methods of this kind have been used in macroeconometrics for DSGE models \citep{HerbstSchorfheide2014} and for Markov-switching VARs \citep{BognanniHerbst2018}.

\subsubsection{Synthetic Data}

First, we illustrate the behavior of $\text{SMC}^2$ on data simulated as follows. We take $T = 10^3$ time steps, $K = 4$ variables and $p = 4$ lags. The autoregressive matrix $\Pi$ is randomly generated under the constraint that its maximum eigenvalue modulus is strictly less than one. Furthermore, we introduce cross-sectional dependence by setting the non-zero off-diagonal elements of matrix $A$ to $0.3$. We set the volatility innovation variance to $\Phi_{j,j} = 0.001$ to reflect smooth, persistent volatility dynamics, also preventing particle degeneracy of the algorithm.

The initial $\theta$-particles are drawn from the particle Gibbs sampler of Section \ref{sec:ourMCMC} run on $y_{1:t_0}$, $t_0=360$. (In other words, we
use the posterior $\pi(\theta|y_{0:t_0})$ as the prior in SMC$^2$.)  The sequential phase runs from $t_0+1$ to $T$ with $N_x = 300$ inner and $N_\theta=7,500$ outer particles, $N_x$ held fixed throughout. Each period $y_t$ is assimilated in $K=4$ univariate steps (Section \ref{sub:smc2:recursive}), so the run has $K(T-t_0)=2560$ steps in total. Rejuvenation is triggered when the ESS falls below $0.5 \times N_\theta = 3,750$ upon observing a full month of data, and below $0.1 \times N_\theta = 750$ during the intermediate steps. The later is set to prevent over-rejuvenating on partial observed data. 

\begin{figure}[htbp]
    \centering
    \includegraphics[width=0.9\linewidth]{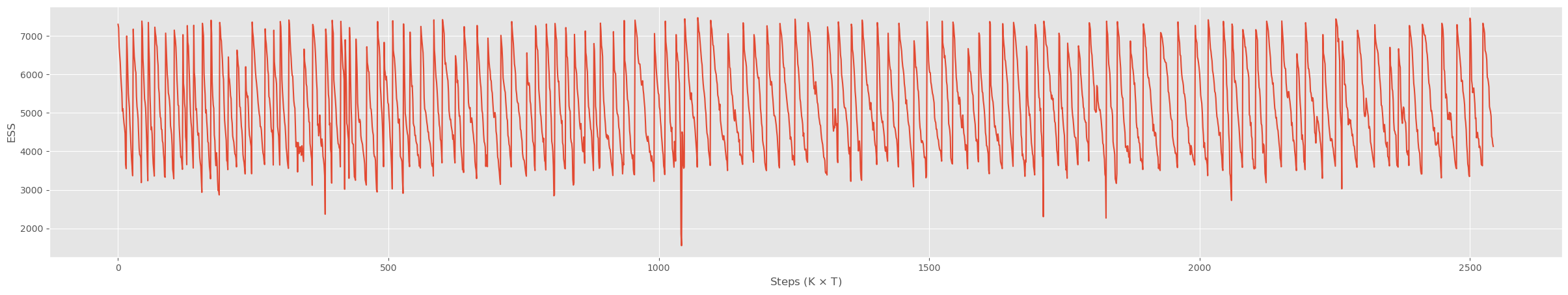}
    \caption{ESS of the $\theta$-particles in one $\text{SMC}^2$ run on synthetic data ($K=4$, $p=4$, $T=1000$, $t_0=360$, $N_\theta=7500$, $N_x=300$). One step on the horizontal axis is one univariate assimilation step, with $K=4$ steps per period.}\label{fig:ess_syn}
\end{figure}

Figure \ref{fig:ess_syn} displays the ESS trace of one $\text{SMC}^2$ run. After each rejuvenation the particles are resampled and the ESS returns to $N_\theta=7500$. It then decays until it hits the threshold. The ESS drops sharply in the very first steps and stabilises afterwards. Rejuvenations are most frequent at the start of the run and their spacing grows only slightly afterwards. This is expected. As $t$ grows, the posterior concentrates and the weights degrade more slowly, but $t$ only moves from 360 to 1000 here, so the information per new observation falls by less than a factor of three over the run. 

On average, rejuvenation occurs every 5 periods, that is, every $20$ algorithm steps. This shows the algorithm is efficient in the regime where $K$ is small
and $T$ is large.

\begin{figure}[htbp]
    \centering
    \includegraphics[width=0.9\linewidth]{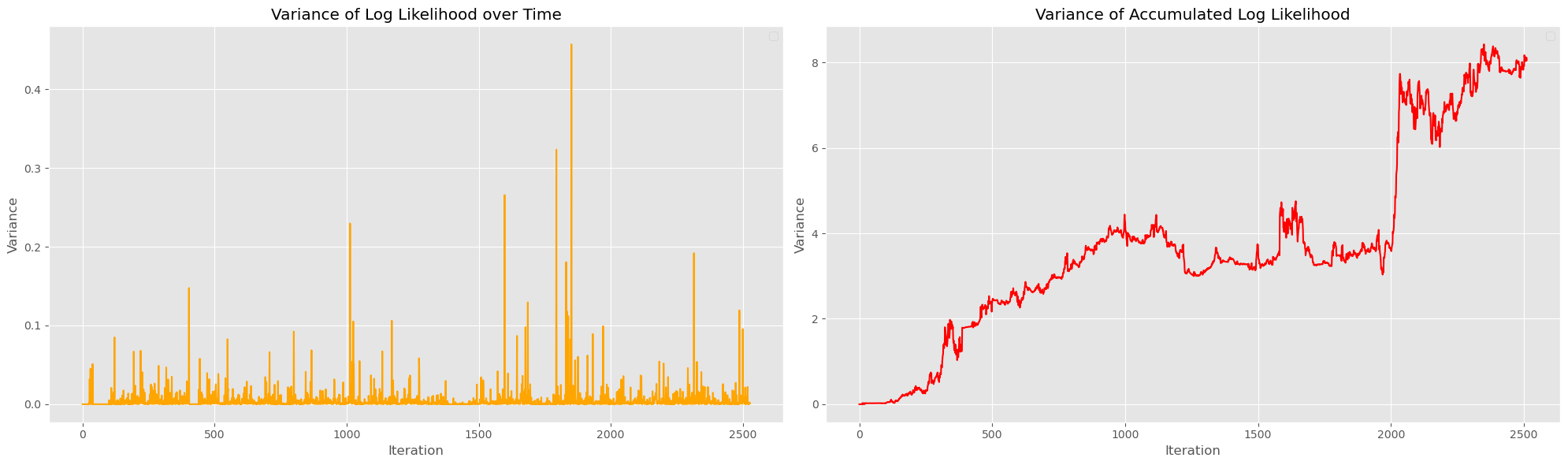}
    \caption{Variance across 20 runs of $\text{SMC}^2$ on the same synthetic dataset. Left panel: variance of the one-step log-likelihood estimates. Right panel: variance of the accumulated log-likelihood. The horizontal axis counts univariate assimilation steps.}
    \label{fig:var_syn}
\end{figure}

The $\text{SMC}^2$ algorithm also provides, at each $t > t_0$, an unbiased estimate $\hat{p}(y_{t_0+1:t} \mid y_{1:t_0})$ of the predictive likelihood of $y_{t_0+1:t}$ given the training sample, that is, the marginal likelihood of the post-training data with the parameters integrated against their posterior given $y_{1:t_0}$. Its logarithm, $\log\hat{p}(y_{t_0+1:t} \mid y_{1:t_0}) = \sum_{s = t_0+1}^{t} \log \hat{p}(y_s \mid y_{1:s-1})$, can be used for model selection. We run 20 independent realizations of $\text{SMC}^2$ on the same simulated dataset and compute, for each $s$ with $t_0 < s \leq T$, the variance across runs of $\log\hat{p}(y_{t_0+1:s} \mid y_{1:t_0})$, as shown in Figure \ref{fig:var_syn}. The left panel reports the variance of the one-step estimates. It is small and spiky, mostly below $0.05$. The right panel reports the variance of the accumulated log-likelihood, which grows to about $8$ by the end of the run. It behaves roughly like the running sum of the left panel, so the estimation errors accumulate slowly from step to step.

%

\subsubsection{Real Data}
To illustrate our \smcsq algorithm in a realistic setting, we use it to select among competing VAR specifications on macroeconomic data. We take $K = 6$ series from Table \ref{tab:pg_data_set} (INDPRO, UNRATE, PCEPI, HOUST, BAAFFM, and GS5), with the same transformations as before. The six series cover real activity, the labor market, prices, housing, interest rates, and credit conditions. They enter in the order listed, which fixes the recursive structure of $A$. The Cholesky specification is not invariant to this order, so each ordering defines a different model and a different marginal likelihood. The sample runs from 1980:M1 to 2021:M3 (495 raw observations). We split the sample at the end of 1999. The training sample is 1980:M1 to 1999:M12, that is, $t_0 = 240$ observations, and each model conditions on its own $p$ initial lags within this window. The $\theta$-particles are initialised with draws from the particle Gibbs sampler run on the training sample. The predictive likelihood is then accumulated over the same prediction sample for all models, 2000:M1 to 2021:M3. The algorithm settings are the same as in the synthetic experiment. Each period is now assimilated in $K=6$ univariate steps rather than $4$, with $N_x$ unchanged.

We compare heteroskedastic VARs with lag lengths $p \in \{2, 4, 6\}$ against a homoskedastic VAR with $p = 2$. The homoskedastic benchmark uses the same algorithm, with the prior on the volatility innovation variances concentrated near zero so that the volatility paths are nearly constant over time. Specifically, we impose $\Phi_{k,k} \sim \mathcal{I}\Gamma\left(25000,1/2\right)$, for $k\in [K]$, which results in expectation of order $10^{-5}$. This nests the homoskedastic model only approximately, but it keeps the implementation identical across all models.

\begin{figure}
    \centering
    \includegraphics[width=0.75\linewidth]{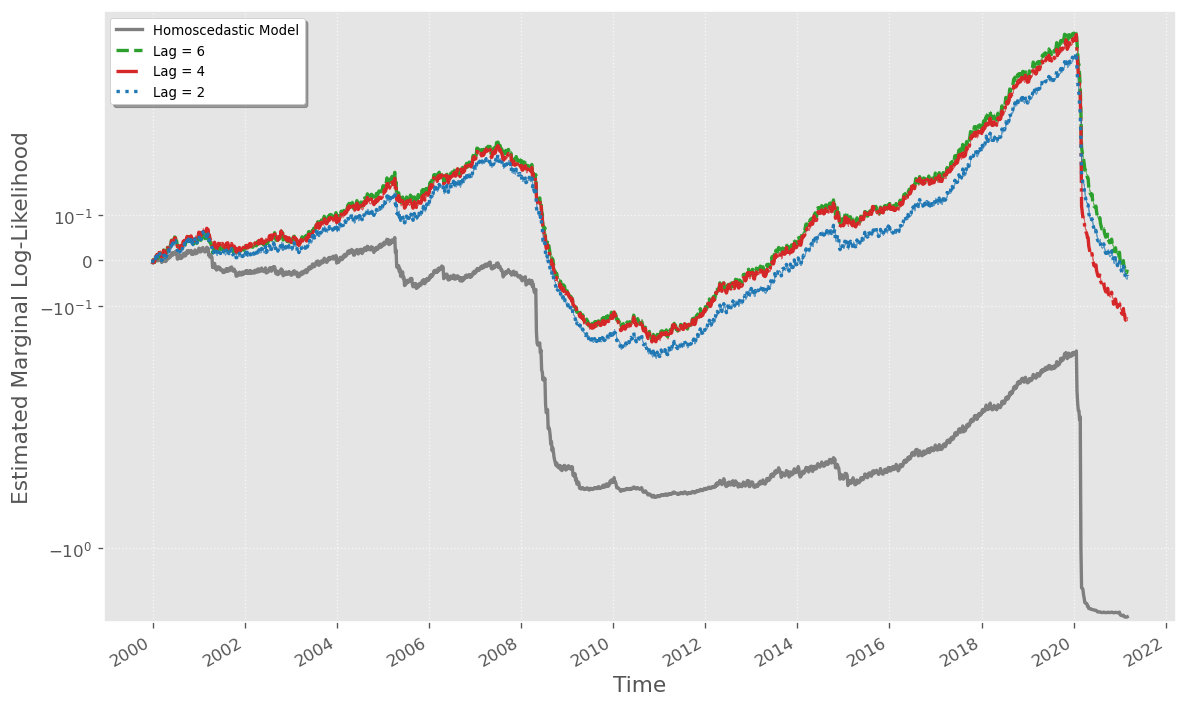}
    \caption{Conditional marginal log-likelihood (conditional on the training sample) $\log\hat{p}(y_{t_0+1:t} \mid y_{1:t_0})$ over time, shown on a symmetric log scale, for heteroskedastic VARs with $p \in \{2,4,6\}$ and the homoskedastic benchmark. Real data, $K=6$, prediction sample 2000:M1 to 2021:M3.}\label{fig:marginal_log_like}
\end{figure}

Figure \ref{fig:marginal_log_like} shows the cumulative log predictive likelihood over time. All models show steep decreases in 2008 and in 2020, when the one-step predictive densities are low during the financial crisis and the COVID-19 pandemic. The heteroskedastic models attain higher predictive likelihood than the homoskedastic benchmark, and the gap widens precisely in these episodes. This matches the evidence that stochastic volatility improves density forecasts of macroeconomic data, with the largest gains in volatile periods \citep{Clark2011, ClarkRavazzolo2015}. At the end of the sample, the gap is about $2.17$ log points. Among the heteroskedastic models the differences are small, within $0.10$ log points, and the ranking of the lag lengths changes with the pandemic observations.

\section{Conclusion}\label{sec:conclusion}

This paper makes two contributions to Bayesian inference for large VAR models with
Cholesky stochastic volatility. The first is an MCMC kernel that targets the
exact posterior, without the mixture approximation of
\citet{KimShephardChib1998} or the associated ordering constraints of
\citet{DelNegroPrimiceri2015}. It rests on a reparametrisation that decouples
the $K$ equations, so that the coefficients are updated row by row and the
volatility paths by $K$ parallel CSMC steps. In an application with $K = 15$
monthly series and $p = 9$ lags, the kernel raises the effective sample size
per second by a factor of about 14 for the VAR coefficients and about 3.4 for
the volatilities relative to the corrected triangular algorithm, at the same
computational complexity. 

The second contribution is a waste-free \smcsq{} sampler built on the same
particle filter. It processes the data one equation at a time, handles
partially observed dates by truncation rather than imputation, and delivers,
on-line, the posterior, the one-step-ahead predictive density and the marginal
likelihood. On real data with $K = 6$, it tracks the relative predictive
likelihood of competing lag lengths and of the homoskedastic benchmark over two
decades, including the 2008 and 2020 episodes.

\section*{Acknowledgements}

The authors acknowledge funding from ANR grant BLISS grant number ANR-24-CE26-1611-01. Anna Simoni gratefully
acknowledges financial support from ANR-21-CE26-0003. This work benefited from Hi! PARIS and State funding managed by the French National Research Agency
(ANR) under the France 2030 program, reference ANR-23-IACL-0005. Yuedan Huo acknowledges a PhD fund granted by FMJH. We are grateful to Pierre Jacob for
sharing with us the code implementing the genealogy trick of \citet{pathstorage}. 

\begin{singlespace}
\putbib
\end{singlespace}
\end{bibunit}

\newpage
\appendix
\section*{Appendix}
\begin{bibunit}
\section{Additional Figure for the numerical experiments of Section \ref{sec:num}}\label{App:figures}

\begin{figure}
    \centering
    \includegraphics[width=0.8\linewidth]{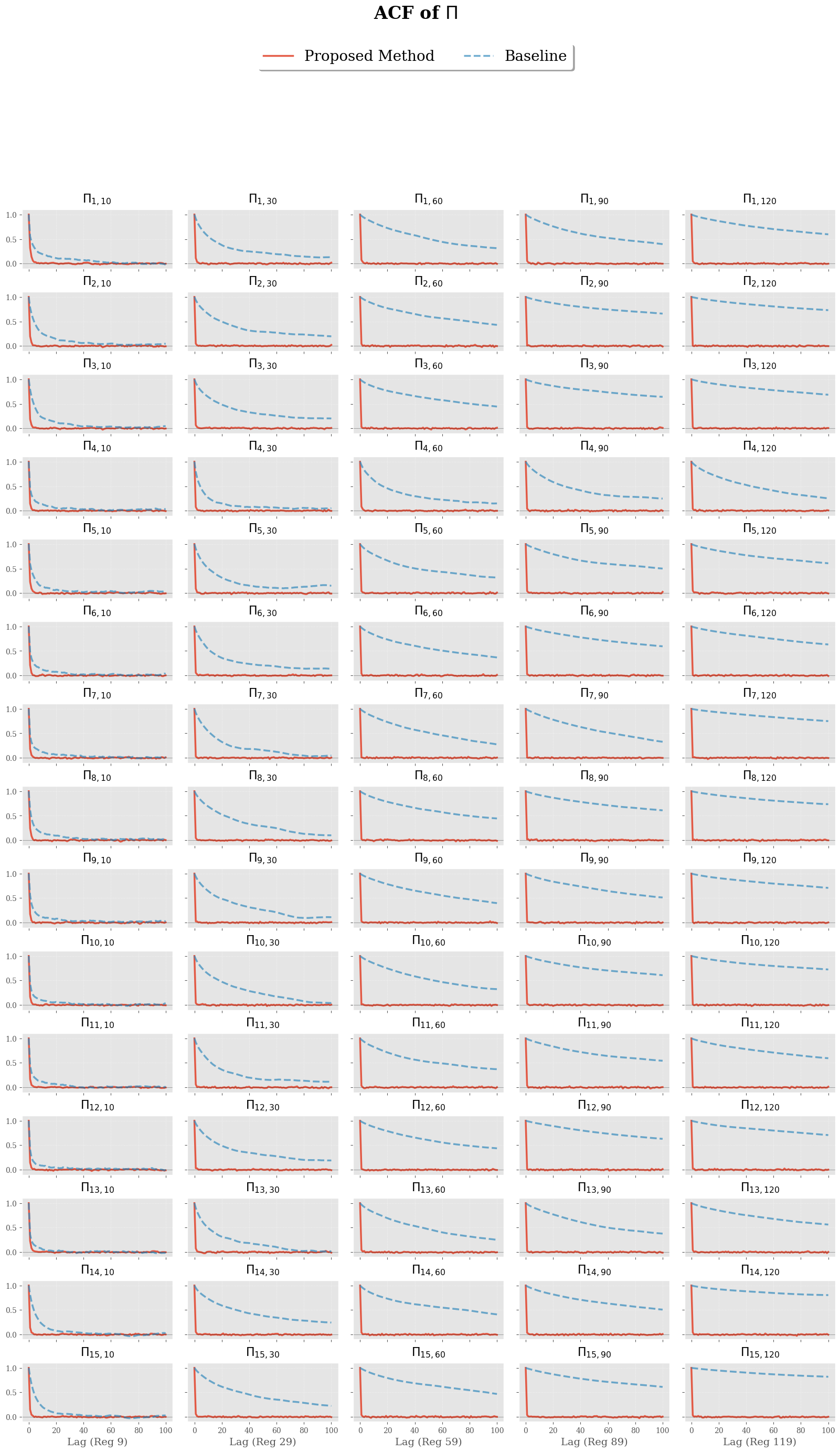}
    \caption{Sample autocorrelation functions for an extended selection of elements of $\Pi$, under the reparametrised update ($\wtl{\Pi}=A\Pi$) and the CTA baseline. It complements Figure~\ref{fig:pi_mixing}. Same data and sampler settings as in Section~\ref{sec:experiment mcmc}.}
    \label{fig:ACF:Pi}
\end{figure}

\begin{figure}
    \centering
    \includegraphics[height=0.88\textheight, width=\linewidth, keepaspectratio]{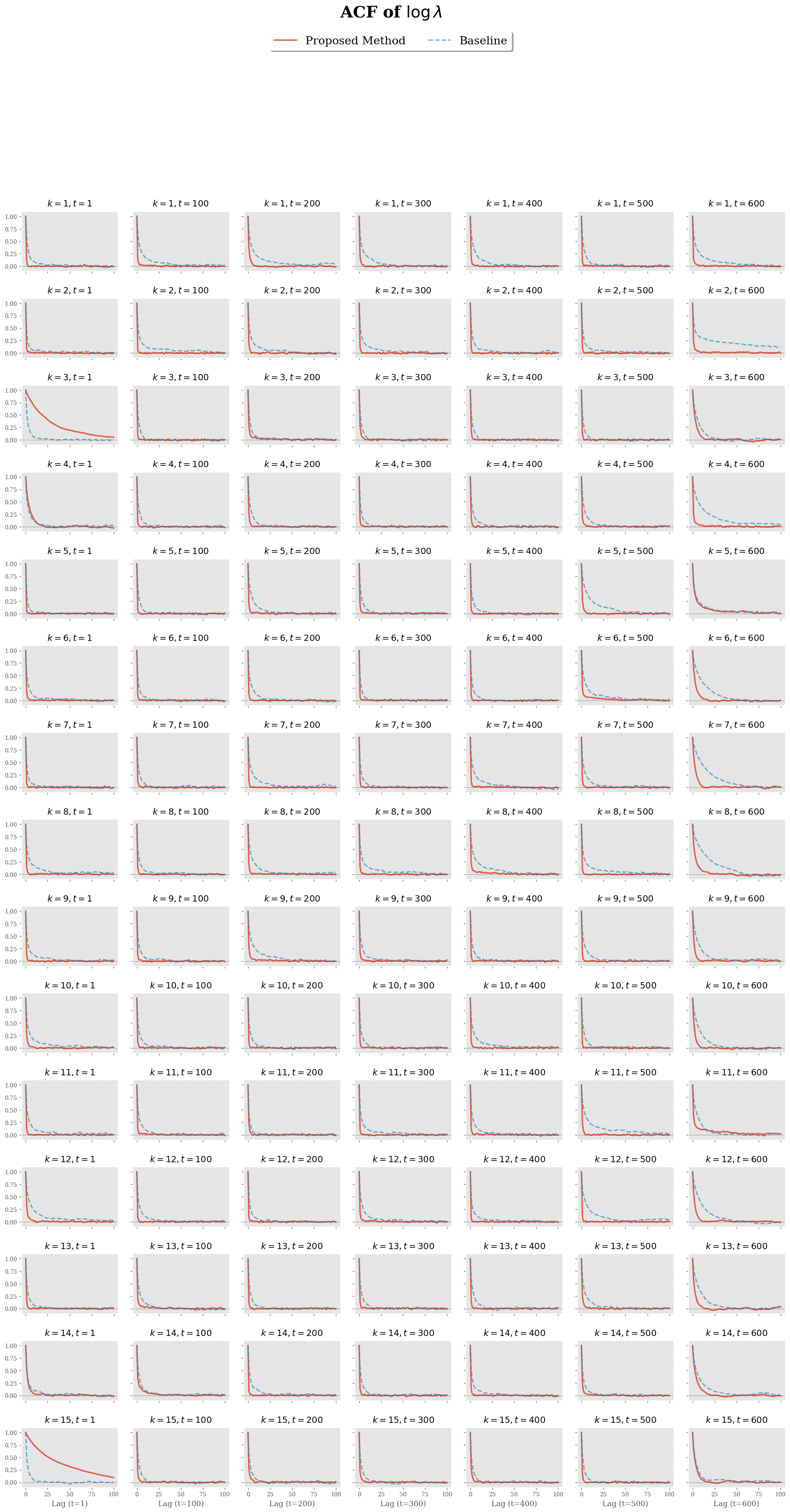}
    \caption{Sample autocorrelation functions for the log-volatility states $\log\lambda_{t,k}$, for all $K=15$ series (rows) at periods $t \in \{1, 100, 200, 300, 400, 500, 600\}$ (columns), under the CSMC update and the KSC mixture sampler. Complements Figure \ref{fig:loglambda_mixing}. Same data and sampler settings as in Section \ref{sec:experiment mcmc}.}
    \label{fig:ACF:log:lambda}
\end{figure}

\section{The PMMH sampler}\label{app:pmmh}
For completeness, we spell out the PMMH (particle marginal Metropolis-Hastings) sampler of \citet{PMCMC} for the univariate stochastic volatility model \eqref{eq:basicSV} of Section \ref{sec:pmcmc}, where the parameter is $\theta = \sigma^2$. We stress that we do not use PMMH in this paper. We describe it only to give the reader some high-level intuition and to contrast it with the CSMC-based approach that we adopt (see Section \ref{sec:ourMCMC}).

The idea is simply to run a Metropolis-Hastings sampler on $\theta$, for instance with a random-walk proposal, targeting the marginal posterior $\pi(\theta \mid y_{1:T}) \propto p_\theta(y_{1:T})\,\pi(\theta)$. The only change relative to the standard algorithm concerns the likelihood $p_\theta(y_{1:T})$, which is intractable here. Whenever a new value of $\theta$ is proposed, the likelihood at that value is replaced by the estimate $\wh{p}_\theta(y_{1:T}) = \prod_{t=1}^T \bigl( \tfrac{1}{N} \sum_{n=1}^N w_t^n \bigr)$, computed from a single run of the particle filter (Algorithm \ref{alg:bootstrap}) at that value, where the $w_t^n$ are the weights produced by the filter. The estimate attached to the current value is stored and reused in the acceptance ratio, not recomputed. The remarkable fact established by \citet{PMCMC} is that this substitution does not change the target: the chain still has invariant distribution $\pi(\theta \mid y_{1:T})$, exactly, and for any fixed $N \geq 1$. The reason is that the pair formed by $\theta$ and the random variables generated by its particle filter is an ordinary Metropolis-Hastings chain on an extended space, and the $\theta$-marginal of the extended target is the exact posterior, because the estimate is unbiased and nonnegative. The number of particles $N$ therefore affects only the mixing of the chain. As $N \to \infty$ the estimate concentrates on $p_\theta(y_{1:T})$ and PMMH behaves like an idealised Metropolis-Hastings sampler run on the exact likelihood.

\begin{algorithm}
  \caption{PMMH (particle marginal Metropolis--Hastings) for the univariate SV model}\label{alg:pmmh}
  \begin{algorithmic}
    \Require data $y_{1:T}$, prior $\pi(\theta)$, proposal $q(\cdot\mid\theta)$, number of particles $N$, number of MCMC iterations $M$ (a local symbol, distinct from the number of waste-free chains of Section \ref{sec:SMC2})
    \State Set $\theta(0)$ arbitrarily
    \State Run Algorithm~\ref{alg:bootstrap} at $\theta(0)$ to obtain
      $\wh{p}_{\theta(0)}(y_{1:T})$
    \For{$i=1,\dots,M$}
      \State $\theta^\star \sim q(\cdot\mid\theta(i-1))$
      \hfill\Comment{Propose a new parameter}
      \State Run Algorithm~\ref{alg:bootstrap} at $\theta^\star$ to obtain
        $\wh{p}_{\theta^\star}(y_{1:T})$
      \hfill\Comment{Unbiased likelihood estimate}
      \State $\alpha \gets 1 \wedge
        \dfrac{\wh{p}_{\theta^\star}(y_{1:T})\,\pi(\theta^\star)\,
               q(\theta(i-1)\mid\theta^\star)}
              {\wh{p}_{\theta(i-1)}(y_{1:T})\,\pi(\theta(i-1))\,
               q(\theta^\star\mid\theta(i-1))}$
      \State With probability $\alpha$, set $\theta(i)\gets\theta^\star$ and $\wh{p}_{\theta(i)}(y_{1:T})\gets\wh{p}_{\theta^\star}(y_{1:T})$
      \Statex \hspace{\algorithmicindent} otherwise set
        $\theta(i)\gets\theta(i-1)$ and
        $\wh{p}_{\theta(i)}(y_{1:T})\gets\wh{p}_{\theta(i-1)}(y_{1:T})$
        \hfill\Comment{Accept or reject}
    \EndFor\\
    
    \Return $\theta(1),\dots,\theta(M)$
  \end{algorithmic}
\end{algorithm}

The main practical drawbacks of PMMH, which motivate our alternative, are the following. First, it explores $\theta$ through a proposal distribution, typically a random walk, whose calibration is notoriously difficult and which scales poorly with the dimension of $\theta$. Second, it re-runs an entire particle filter at every iteration. See the discussion in Section~\ref{sec:pmcmc} and \citet[][Chap. 16]{SMCbook}.

A further drawback concerns the choice of $N$. Although PMMH is exact for any fixed $N \geq 1$, its efficiency hinges on the variance $\sigma^2_N$ of the log-likelihood estimate $\log \wh{p}_\theta(y_{1:T})$, which should be held close to a constant of order one (typical targets are $\sigma^2_N \approx 1$ - $3$, the latter matching the optimal acceptance rate of about $7\%$ derived by \citet{Sherlock2015}, see also \citet{Pitt2012134} and \citet{MR3371005}). Since this variance typically grows linearly in $T$ and increases with the latent-state dimension, one must take $N = \bigO(T)$, and larger still as the state dimension grows, merely to keep the acceptance rate from collapsing. This makes PMMH far more costly than the conditional-SMC approach of Section \ref{sec:ourMCMC} for our $K$-dimensional log-volatility state and samples of several hundred observations.

\section{Prior details}\label{app:prior}

\subsection{Minnesota prior: hyperparameters}\label{app:minnesota}

This appendix specifies the Minnesota hyperparameters used in Section \ref{sec:model}. For the intercepts,
\begin{eqnarray*}
  \Pi_0 & \sim & \mathcal{N}_K(0_K, 100 \,I_K),
\end{eqnarray*}
\noindent while, for the autoregressive coefficients, each entry has an independent Gaussian prior, mutually independent across equations $k$, predictors $i$, and lags $j$ (and independent of $\Pi_0$):
\begin{displaymath}
  (\Pi_j)_{k,i} \sim \mathcal{N}\!\left((\mu_j)_{k,i},\, \operatorname{Var}\!\left((\Pi_j)_{k,i}\right)\right), \qquad j=1,\ldots,p,\quad k,i\in[K].
\end{displaymath}
\noindent The means and variances are specified as in the Minnesota prior originally proposed by \citet{Litterman1979,Litterman1986} and refined by \citet{KadiyalaKarlsson1997}, \citet{SimsZha1998}, and \citet{BanburaGiannoneReichlin2010} among others: for $j=1,\ldots,p$
\begin{displaymath}
  (\mu_j)_{k,i} = \left\{\begin{array}{cc}
    \delta_i, & k=i,\,j=1\\
    0, & \textrm{ otherwise}
  \end{array}\right., \qquad \operatorname{Var}((\Pi_j)_{k,i}) = \left\{\begin{array}{cc}
    \frac{\rho_1}{j^{\rho_3}}, & k=i\\
    \rho_2\frac{\rho_1}{j^{\rho_3}}\frac{\sigma_k^2}{\sigma_i^2}, & \textrm{ otherwise}
  \end{array}\right.,
\end{displaymath}
\noindent where $\rho \coloneq (\rho_1,\rho_2,\rho_3)^\top$ is a vector of hyperparameters to be specified and $\sigma_i^2$ is the scale parameter which we set equal to the variance of the residuals from a univariate autoregressive model with drift for the $i$-th series, as it is standard in the literature (\textit{e.g.} \citet{Litterman1986}, \cite{SimsZha1998}, and \citet{CARRIERO2019137}). Note that for own lags $(k = i)$ the ratio $\sigma_k^2/\sigma_i^2$ equals one, so the two parts of the variance differ only by the cross-variable factor $\rho_2$. We refer to \citet{BanburaGiannoneReichlin2010} for a detailed  description of the role played by these hyperparameters. The value $\delta_i$ is set to $1$ if the $i$-th series is persistent, and to $0$ if it exhibits mean reversion (implying a white noise prior is more appropriate).  The lag-1 autocorrelation is computed as
\begin{equation*}
    \hat{\rho}_1 = \frac{\sum_{t=1}^{T-1}(y_t - \bar{y})(y_{t+1} - \bar{y})}{\sum_{t=1}^{T}(y_t - \bar{y})^2},
\end{equation*}
where $y_t$ is simplified for $y_{t,i}$ as it is computed per series. Series with $\hat{\rho}_1 > 0.5$ are classified as persistent while series with $\hat{\rho}_1 \leq 0.5$ are classified as not persistent. See Table \ref{tab:variable_persistence} for the classification of the variables in our application.
\begin{table}[htbp]
  \centering
  \begin{tabular}{r l r l}
    \toprule
    \textbf{\#} & \textbf{Variable} & \textbf{Lag-1 corr} & \textbf{Classification} \\
    \midrule
     0 & RPI             & $-0.4738$ & not persistent \\
     1 & INDPRO          & $0.2834$  & not persistent \\
     2 & CUMFNS          & $0.9827$  & \textbf{persistent} \\
     3 & UNRATE          & $0.9589$  & \textbf{persistent} \\
     4 & PAYEMS          & $0.0589$  & not persistent \\
     5 & CES0600000007   & $0.9093$  & \textbf{persistent} \\
     6 & CES0600000008   & $0.2443$  & not persistent \\
     7 & PCEPI           & $0.7144$  & \textbf{persistent} \\
     8 & WPSFD49207      & $0.3267$  & not persistent \\
     9 & HOUST           & $0.9697$  & \textbf{persistent} \\
    10 & DPCERA3M086SBEA  & $0.0167$  & not persistent \\
    11 & EXUSUK          & $0.3342$  & not persistent \\
    12 & BAAFFM          & $0.9668$  & \textbf{persistent} \\
    13 & GS5             & $0.9949$  & \textbf{persistent} \\
    14 & GS10            & $0.9955$  & \textbf{persistent} \\
    \bottomrule
  \end{tabular}
  \caption{Lag-1 autocorrelation and persistence classification (Threshold = $0.5$).}
  \label{tab:variable_persistence}
\end{table}
For a medium-sized model, we set the parameters as follows: $\rho_1 = 0.05$, $\rho_2 = 0.5$ and $\rho_3 = 2$, with $\rho_1 = 0.05$ matching the choice in \citet{CARRIERO2019137} for $N=20$.

The Minnesota prior is motivated by the following intuition. Rather than centering every equation on the same dynamics, it shrinks each variable toward a univariate benchmark whose persistence is governed by $\delta_i$. Concretely, the prior shrinks the own-first-lag coefficient $(\Pi_1)_{i,i}$ toward $\delta_i$ and all remaining coefficients in $\Pi_1,\ldots,\Pi_p$ toward zero. When $\delta_i = 1$, the $i$-th equation is centered on the random walk with drift $y_{i,t} = \Pi_{0,i} + y_{i,t-1} + v_{i,t}$, encouraging persistent, random-walk behaviour. When $\delta_i = 0$, it is instead centered on a white-noise process $y_{i,t} = \Pi_{0,i} + v_{i,t}$, which is appropriate for mean-reverting series. The prior variance reflects two key beliefs: (1) more recent lags are more informative than distant lags, and (2) a variable's own lags are more relevant than other variables' lags.

\subsection{Derivation of the conditional prior for \texorpdfstring{$\wtl{\Pi}_{(j)}$}{Pi tilde (j)}}\label{app:priorderiv}
This appendix derives the Gaussian conditional prior $\pi(\wtl{\Pi}_{(j)}\mid\wtl{\Pi}_{(-j)},A)$ whose precision \eqref{eq:prior:precision} and mean \eqref{eq:prior:mean} are stated in Section \ref{sec:model}. Recall that $L \coloneq A^{-1}$ is unit lower triangular and that $\wtl\Pi = A\Pi$  (see \eqref{model:eq:reparametrized}) so the rows of $\Pi = L\wtl{\Pi}$ satisfy
\begin{equation}\label{eq_Pi_k}
  \Pi_{(k)} = \sum_{m=1}^{k} L_{k,m} \wtl{\Pi}_{(m)} = \wtl{\Pi}_{(k)} + \sum_{m=1}^{k-1} L_{k,m} \wtl{\Pi}_{(m)}, \qquad k \in [K],
\end{equation}
using $L_{k,k} = 1$. In particular, $\wtl{\Pi}_{(j)}$ enters $\Pi_{(k)}$ only for $k \geq j$. Since the map $\wtl \Pi \mapsto L\wtl{\Pi}$ is linear with
$|L|^{Kp+1} = 1$, the prior of Section \ref{sec:model} induces, by a change of variables without Jacobian factor, the following log-density for $\wtl{\Pi}$ given $A$, up to an additive constant:
\begin{eqnarray}\label{eq:logprior_Pitilde}
  \ln \pi(\wtl{\Pi} \mid A) & = & -\frac{1}{2} \sum_{k=1}^{K}\left(\sum_{m=1}^{k} L_{k,m}\wtl{\Pi}_{(m)} - \mu_{\Pi_{(k)}}\right)^{\!\top}\Sigma_{\Pi_{(k)}}^{-1}\left(\sum_{m=1}^{k} L_{k,m}\wtl{\Pi}_{(m)} - \mu_{\Pi_{(k)}}\right) + \text{const}\nonumber\\
  & = & -\frac{1}{2} \sum_{k=1}^{K} (\Pi_{(k)} - \mu_{\Pi_{(k)}})^\top \Sigma_{\Pi_{(k)}}^{-1} (\Pi_{(k)} - \mu_{\Pi_{(k)}}) + \text{const}.
\end{eqnarray}
%
\noindent To derive the density for $\pi(\wtl{\Pi}_{(j)} \mid \wtl{\Pi}_{(-j)},
A)$, we isolate all terms in the sum where index $\wtl\Pi_j$ appears. By
\eqref{eq_Pi_k}, these are the summands with $k\geq j$. 
\noindent Denote by $\ell_k(\wtl{\Pi})$ the $k$-th summand of $\ln\pi(\wtl{\Pi})$. For each $k\geq j$, we isolate the term $m=j$ in \eqref{eq_Pi_k}: 
\begin{multline*}
    \ell_k(\wtl{\Pi}) \propto -\frac{1}{2}\left(\sum_{m\leq
    k}L_{k,m}\wtl{\Pi}_{(m)} - \mu_{\Pi_{(k)}}\right)^\top \Sigma_{\Pi_{(k)}}^{-1}\left(\sum_{m\leq k}L_{k,m}\wtl{\Pi}_{(m)} - \mu_{\Pi_{(k)}}\right) \\
    \propto -\frac{1}{2}\left(L_{k,j}\wtl{\Pi}_{(j)} + \sum_{m\leq k, m\neq
    j}L_{k,m}\wtl{\Pi}_{(m)} - \mu_{\Pi_{(k)}}\right)^\top \Sigma_{\Pi_{(k)}}^{-1} \left(L_{k,j}\wtl{\Pi}_{(j)} + \sum_{m\leq k, m\neq j}L_{k,m}\wtl{\Pi}_{(m)} - \mu_{\Pi_{(k)}}\right).
\end{multline*}
\noindent Because $L$ is unit lower triangular, $L_{k,j} = 0$ for $k<j$, so the summands $\ell_k(\wtl\Pi)$ with $k<j$ do not involve $\wtl\Pi_{(j)}$ and are absorbed into the normalising constant of the conditional. 
Define the partial prior residual $\mathbf{r}_{k, -j}$ as the part of the prior mean of $\Pi_{(k)}$ that cannot be explained by the other transformed rows $\{\wtl{\Pi}_{(m)}\}_{m\leq k, m\neq j}$, \textit{i.e.},
$$ \mathbf{r}_{k, -j} \coloneq \mu_{\Pi_{(k)}} - \sum_{m \leq k, m \neq j} L_{k,m} \wtl{\Pi}_{(m)}. $$
\noindent It is the residual after accounting for the contributions of all rows except the $j$-th, and, gathered across $k$, it produces the conditional prior mean \eqref{eq:prior:mean}. Importantly, $\mathbf{r}_{k, -j}$ isolates the prior mean component that is independent of $\wtl{\Pi}_{(j)}$, allowing us to compute the conditional prior mean for $\wtl{\Pi}_{(j)}$ without double-counting its contributions. Hence, we can write:
\begin{equation}\label{eq:ellk}
  \ell_k(\wtl{\Pi})
  = -\frac{1}{2}\left(L_{k,j}\wtl{\Pi}_{(j)} - \mathbf{r}_{k,-j}\right)^\top \Sigma_{\Pi_{(k)}}^{-1} \left(L_{k,j}\wtl{\Pi}_{(j)} - \mathbf{r}_{k,-j}\right) + \text{const}.
\end{equation}
\noindent Expanding the quadratic form in \eqref{eq:ellk} and discarding the term $\mathbf{r}_{k,-j}^\top \Sigma_{\Pi_{(k)}}^{-1} \mathbf{r}_{k,-j}$, which is constant in $\wtl{\Pi}_{(j)}$, we obtain
\begin{equation*}
  \ell_k(\wtl{\Pi}) = -\frac{1}{2}\, \wtl{\Pi}_{(j)}^\top \left(L_{k,j}^2\, \Sigma_{\Pi_{(k)}}^{-1}\right) \wtl{\Pi}_{(j)} + \wtl{\Pi}_{(j)}^\top    \left(L_{k,j}\, \Sigma_{\Pi_{(k)}}^{-1}\, \mathbf{r}_{k,-j}\right) + \text{const}.
\end{equation*}
\noindent Summing over $k \geq j$ and using $L_{j,j} = 1$, the log conditional prior is, up to an additive constant,
\begin{equation}\label{eq:logcondprior}
  \ln \pi\!\left(\wtl{\Pi}_{(j)} \,\middle|\, \wtl{\Pi}_{(-j)}, A\right) = -\frac{1}{2}\, \wtl{\Pi}_{(j)}^\top\, Q_j\, \wtl{\Pi}_{(j)} + \wtl{\Pi}_{(j)}^\top\, b_j + \text{const},
\end{equation}
with
\begin{equation*}
  Q_j \coloneq \Sigma_{\Pi_{(j)}}^{-1}
      + \sum_{k=j+1}^{K} L_{k,j}^2\, \Sigma_{\Pi_{(k)}}^{-1},
  \qquad
  b_j \coloneq \Sigma_{\Pi_{(j)}}^{-1}\mathbf{r}_{j,-j}
      + \sum_{k=j+1}^{K} L_{k,j}\, \Sigma_{\Pi_{(k)}}^{-1}\, \mathbf{r}_{k,-j},
\end{equation*}
where the $k=j$ contributions are written separately using $L_{j,j}=1$. Here the term $\Sigma_{\Pi_{(j)}}^{-1}$ comes from the prior on $\Pi_{(j)}$ and the terms $L_{k,j}^2 \Sigma_{\Pi_{(k)}}^{-1}$ come from the contribution of $\wtl{\Pi}_{(j)}$ to $\Pi_{(k)}$ for $k > j$, due to the lower triangular structure of $L$. Since \eqref{eq:logcondprior} is a quadratic form in $\wtl{\Pi}_{(j)}$ with positive-definite $Q_j$ (each $\Sigma_{\Pi_{(k)}}$ is positive definite and
the additional terms are positive semi-definite), completing the square yields
\begin{equation*}
  \wtl{\Pi}_{(j)} \mid \wtl{\Pi}_{(-j)}, A \sim \mathcal{N}\!\left(\mu_{\mathrm{prior},j}, \Omega_{\mathrm{prior},j}\right),
  \qquad \Omega_{\mathrm{prior},j} = Q_j^{-1}, \quad \mu_{\mathrm{prior},j} = Q_j^{-1} b_j,
\end{equation*}
which are exactly the precision \eqref{eq:prior:precision} and mean
\eqref{eq:prior:mean} stated in Section~\ref{sec:model}.

\section{Comparison with the corrected triangular algorithm (CTA)}\label{app:cta}
This appendix expands Remark \ref{rem:comparison_CTA}. We first account for the computational cost of our update and of the corrected triangular algorithm (CTA) of \citet{CarrieroChanClarkMarcellino2022}, showing that they share the same leading order, and then give the mixing-rate argument that is the substantive difference between them.

Both our update and the CTA are equation-by-equation Gibbs sweeps, and they are in fact closely parallel. Write
\[
  G_j \coloneq \sum_{t=1}^T \frac{z_t z_t^\top}{\lambda_{t,j}}
\]
for the volatility-weighted Gram matrix that appears as the likelihood precision in the full conditional of Section~\ref{sec:ourMCMC}. The CTA draws the reduced-form rows $\Pi_{(j)}$ directly. Because the reduced-form residual $v_{t,j} = y_{t,j} - \Pi_{(j)}^\top z_t$ enters the structural shock of every downstream equation $m \geq j$, the correct full conditional augments equation $j$ with those downstream equations, and its precision is
\begin{equation*}
  M_j^{\mathrm{CTA}} = \underbrace{\sum_{m=j}^{K} A_{mj}^2\, G_m}_{\text{coupling in the likelihood}} + \Sigma_{\Pi_{(j)}}^{-1},
\end{equation*}
where $A_{jj} = 1$, so that the term $m = j$ is the own-equation Gram $G_j$. The cross-equation sum sits in the likelihood, while the prior stays block-diagonal. Our update draws instead the rows of the reparametrised matrix, $\wtl{\Pi}_{(j)}^\top = A_{(j)}^\top \Pi$. The transformed likelihood factorises over rows, so its contribution to the precision of row $j$ is $G_j$ alone, and an analogous cross-equation sum appears in the prior instead, now running over the entries of $L = A^{-1}$ and the prior precisions:
\begin{equation*}
  \bar{\Sigma}_{\wtl{\Pi}_{(j)}}^{-1} = G_j + \underbrace{\Big(\Sigma_{\Pi_{(j)}}^{-1} + \sum_{k=j+1}^{K} L_{kj}^2\,\Sigma_{\Pi_{(k)}}^{-1}\Big)}_{\text{coupling in the prior}},
\end{equation*}
as in \eqref{eq:prior:precision}. The two sums are of different magnitudes: the likelihood-borne coupling of the CTA involves the Grams $G_m$, which grow with $T$, whereas the prior-borne coupling of our update involves fixed prior precisions. The placement and size of the cross-equation coupling is the heart of the comparison. The two updates cost essentially the same to assemble, as we show next, and it is where the coupling lives that changes the mixing.

Both updates admit implementations of the same leading order once the moments are cached, and it is in that form that they should be compared. The code released with the corrigendum rebuilds, for each $j$, a stacked design of $T(K-j+1)$ rows, at a cost of $\bigO(TK^4p^2)$ per sweep. The extra factor of $K$ is an artefact of that vectorised build rather than of the algorithm, since the corrigendum also states the conditionals in sums-of-moments form. In that form the augmented precision of equation $j$ is the weighted sum $\sum_{m \geq j} A_{mj}^2 G_m$ of the Grams $G_1, \dots, G_K$, and these can be cached, since $z_t$ and $\lambda_{1:T}$ are held fixed while $\Pi$ is drawn. Forming the family $\{G_1, \dots, G_K\}$ once per sweep costs $\bigO(TK^3p^2)$, and this cost is common to both samplers, since our row $j$ uses $G_j$. Beyond it, the CTA pays a further $\bigO(K^4p^2)$ per sweep to recombine the Grams into $\sum_{m \geq j} A_{mj}^2 G_m$ for each $j$. The mean side calls for one additional cache, because the cross moments of the CTA involve the residuals of the downstream equations, which depend on the current draws $\Pi_{(i)}$ and therefore change within the sweep: storing the fitted values $z_t^\top \Pi_{(i)}$ for all $(t, i)$ and refreshing the single column $i = j$ after block $j$ is drawn keeps the mean side at $\bigO(TK^3p)$ per sweep. The remaining work is dominated, for both samplers, by the $K$ factorisations of $(Kp+1)$-dimensional precisions, at $\bigO(K^4p^3)$.\footnote{Both therefore run at $\bigO(TK^3p^2 + K^4p^3)$ per sweep, which is the fine-grained form of the $\bigO(K^4)$ complexity reported by \citet{CarrieroChanClarkMarcellino2022}, counted in the number of variables alone with $T$ and $p$ held fixed.} The genuine differences between the CTA and our algorithm are therefore statistical, not computational.

\textbf{The main advantage: better mixing.} Since both samplers are equation-by-equation Gibbs sweeps, the relevant comparison is the mixing rate rather than the operation count. Fix $(A, \Phi, \lambda_{1:T})$ and stack the drawn coefficients into a single Gaussian vector. Its conditional precision decomposes as $M = M_{\mathrm{lik}} + M_{\mathrm{prior}}$, where $M_{\mathrm{lik}}$ collects the likelihood contributions and $M_{\mathrm{prior}}$ the prior ones. For a Gaussian target, the convergence rate of a systematic-scan block Gibbs sweep is determined by the partial correlations between blocks. Heuristically, the between-equation autocorrelation is governed by the size of the normalised off-diagonal blocks,
\begin{equation*}
  \rho_{ij}\ \sim\ \frac{\lVert M_{ij}\rVert}{\sqrt{\lVert M_{ii}\rVert\,\lVert M_{jj}\rVert}},
  \qquad i \neq j,
\end{equation*}
where $\lVert \cdot \rVert$ denotes the spectral norm. For our sampler, the coordinates are the rows of $\wtl{\Pi}$. The transformed likelihood is block-diagonal in these coordinates, so the off-diagonal blocks of $M$ come from the prior alone, while each diagonal block carries the Gram $G_j$, of order $T$ under ergodicity of the weighted regressors (see the caveats below):
\begin{equation*}
  M_{ij} = \sum_{k \geq \max(i,j)} L_{ki} L_{kj}\, \Sigma_{\Pi_{(k)}}^{-1} = \bigO(1),
  \qquad
  M_{jj} = \underbrace{G_j}_{\bigO(T)} + \underbrace{\Sigma_{\Pi_{(j)}}^{-1} + \sum_{k > j} L_{kj}^2\, \Sigma_{\Pi_{(k)}}^{-1}}_{\bigO(1)}.
\end{equation*}
Hence $\rho_{ij}^{\mathrm{ours}} \sim \bigO(1)/\sqrt{\bigO(T)\,\bigO(T)} = \bigO(1/T) \to 0$. The rows become asymptotically conditionally independent, and the sweep mixes arbitrarily well as data accumulate. The load-bearing fact is data dominance: an $\bigO(T)$ likelihood swamps the fixed $\bigO(1)$ prior coupling. This makes precise the informal statement of Section~\ref{sec:ourMCMC} that the rows are dependent only through the prior. The CTA instead sweeps the reduced-form rows, whose coupling comes from the likelihood. Since $\Sigma_t^{-1} = A^\top \Lambda_t^{-1} A$, the off-diagonal blocks of its joint precision are
\begin{equation*}
  M_{ij}^{\mathrm{CTA}} = \sum_{m \geq \max(i,j)} A_{mi} A_{mj}\, G_m = \bigO(T),
\end{equation*}
weighted sums of the same Grams that make up the diagonal blocks $M_{jj}^{\mathrm{CTA}}$, and therefore of the same order. It follows that
\begin{equation*}
  \rho_{ij}^{\mathrm{CTA}} \sim \frac{-\Sigma_{ij}^{-1}}{\sqrt{\Sigma_{ii}^{-1}\Sigma_{jj}^{-1}}} = \bigO(1),
\end{equation*}
a between-equation autocorrelation that is bounded away from zero regardless of the sample size, and whose magnitude is set by how contemporaneously correlated the variables are, through the sub-diagonal entries of $A$. In the homoskedastic special case, the ratio reduces to the partial correlation between the errors of equations $i$ and $j$. This is consistent with the remark of \citet{CarrieroChanClarkMarcellino2022} that their equation-by-equation blocking ``produces more correlated draws, which might slow down mixing''. The reparametrisation $\wtl{\Pi} = A\Pi$ is precisely what removes this channel.

\textbf{Caveats.} The $\bigO(1/T)$ rate is an asymptotic statement. Since the prior is fixed while $G_j$ grows, the rate always holds, but the size of the constant determines whether the decoupling is felt at realistic sample sizes, and three features of our setting inflate it. First, under tight shrinkage the prior variances are small, which inflates $\Sigma_{\Pi_{(k)}}^{-1}$ in exactly the heavily shrunk directions, namely distant lags and cross-variable coefficients. Second, under strong contemporaneous correlation the off-diagonal entries of $L = A^{-1}$ are large. Third, under weak identification, for instance near-collinear regressors at high lag orders, $G_j$ is nearly singular in some directions, so the likelihood need not dominate the prior in those coordinates. In all three cases the prior coupling remains comparable to the likelihood precision at finite $T$, and the decoupling is correspondingly slower. The decoupling is moreover conditional on $A$. Since $A$, hence $L$, is resampled at each sweep, and $\Pi = L\wtl{\Pi}$ is recovered by back-transformation, strong posterior dependence between $A$ and $\wtl{\Pi}$ is a separate channel that the row decoupling does not address.

\section{Classical Gibbs updates}\label{app:classical}

The MCMC kernel of Section~\ref{sec:ourMCMC} cycles through four blocks. Steps 2 ($\wtl{\Pi}$) and 4 ($\lambda_{1:T}$) are the novel ingredients and are derived in the main text. Steps 1 ($A$) and 3 ($\Phi$) are standard conjugate Gibbs updates, which we record here for completeness.

\subsection{Update of \texorpdfstring{$A$}{A} (Step 1)}\label{app:updateA}
This step is the classical structural-VAR update of \citet{COGLEY2005}. Write the reduced-form residual as $v_t \coloneq y_t - \Pi z_t$, so that \eqref{model:eq:2} reads $A v_t = \Lambda_t^{1/2}\epsilon_t$ with $\epsilon_t\sim\mathcal{N}(0_K, I_K)$, independently across $t$. Because $A$ is lower-triangular with unit diagonal, its $j$-th row satisfies, for every $t\in[T]$,
\begin{equation*}
  A_{(j)}^\top v_t = v_{t,j} + \sum_{k=1}^{j-1} A_{j,k}\, v_{t,k}
  = \lambda_{t,j}^{1/2}\,\epsilon_{t,j}.
\end{equation*}
The unit diagonal also gives $\det A = 1$, so the Gaussian normalising constant of $v_t$ does not depend on $A$ and the conditional density of $v_t$ given $\lambda_t$ is proportional to $\prod_{j=1}^K \exp\{-(A_{(j)}^\top v_t)^2/(2\lambda_{t,j})\}$. Conditionally on $\Pi$ and $\lambda_{1:T}$, the likelihood therefore factorises across the rows of $A$. The prior of Section~\ref{sec:prior} is independent across the free entries, hence across rows, so the full conditional of the free entries factorises as well and the rows may be drawn independently, and in parallel. Only the $j-1$ entries of row $j$ below the diagonal are free. The first row carries none, so $j$ ranges over $2,\dots,K$.

Collect the free entries of row $j$ into the vector $a_{(j)} \coloneq (A_{j,1},\dots,A_{j,j-1})^\top\in\mathbb{R}^{j-1}$ and the corresponding residuals into $v_{t,1:j-1} \coloneq (v_{t,1},\dots,v_{t,j-1})^\top$, so that $A_{(j)}^\top v_t = v_{t,j} + a_{(j)}^\top v_{t,1:j-1}$. Up to a factor that does not depend on $a_{(j)}$, the likelihood of row $j$ is
of row $j$ is
\begin{equation*}
  \prod_{t=1}^T \exp\left\{ -\frac{\bigl(v_{t,j} + a_{(j)}^\top v_{t,1:j-1}\bigr)^2}{2\lambda_{t,j}} \right\},
\end{equation*}
which is proportional to a Gaussian in $a_{(j)}$. Under the Gaussian prior $a_{(j)}\sim\mathcal{N}(\mu_{A,(j)}, \Sigma_{A,(j)})$ specified in Section~\ref{sec:model} (in our case $\mu_{A,(j)}=0_{j-1}$ and $\Sigma_{A,(j)}=10^6\, I_{j-1}$), completing the square yields, for every $j\in\{2,\dots,K\}$,  the Gaussian full conditional
\begin{equation*}
  a_{(j)}\mid y_{1:T}, \Pi, \lambda_{1:T} \sim \mathcal{N}\bigl(\bar{\mu}_{A,(j)}, \bar{\Sigma}_{A,(j)}\bigr),
\end{equation*}
with
\begin{align*}
  \bar{\Sigma}_{A,(j)}^{-1}
  &= \Sigma_{A,(j)}^{-1} + \sum_{t=1}^T \frac{v_{t,1:j-1}\, v_{t,1:j-1}^\top}{\lambda_{t,j}}, \\
  \bar{\mu}_{A,(j)}
  &= \bar{\Sigma}_{A,(j)}\left( \Sigma_{A,(j)}^{-1}\mu_{A,(j)} - \sum_{t=1}^T \frac{v_{t,j}\, v_{t,1:j-1}}{\lambda_{t,j}} \right).
\end{align*}
Neither the likelihood nor the prior of row $j$ involves $\Phi$ or the other rows of $A$, which is why the conditioning set above reduces to $(\Pi, \lambda_{1:T}, y_{1:T})$, as announced in Section~\ref{sec:ourMCMC}. The minus sign in $\bar{\mu}_{A,(j)}$ reflects that $a_{(j)}$ enters the residual additively, so the implied regression is of $-v_{t,j}$ on $v_{t,1:j-1}$. The precision $\bar{\Sigma}_{A,(j)}^{-1}$ is positive definite for every sample size, including $T < j-1$, because $\Sigma_{A,(j)}^{-1}$ is positive definite and the sum is positive semi-definite, so the update never degenerates. After sampling every row, $A$ is reassembled and the transformed data $\wtl{y}_t = A y_t$ are recomputed, as noted in Section~\ref{sec:ourMCMC}.

\subsection{Exact truncation of the \texorpdfstring{$A$}{A}-update under partial observations}\label{app:A_partial}
This appendix justifies the claim of Section~\ref{sub:smc2:rejuv} that, when a resample-move fires in the middle of a period, the rows of $A$ can be rejuvenated by the same per-row truncation used for $\wtl{\Pi}$, running the data sum to $t$ for the observed rows and to $t-1$ for the unobserved ones, with no imputation of the missing block. Suppose that, for some $k\in\{0,\dots,K\}$, the components $y_{t,1:k}$ are observed and $y_{t,k+1:K}$ are not. The periods $y_{1:t-1}$ are complete, so $z_s$ is known for every $s\leq t$. Write the row-$j$ free entries as $a_{(j)}=(A_{j,1},\dots,A_{j,j-1})^\top$. Appendix~\ref{app:updateA} gives the complete-data likelihood contribution of row $j$ at a single period $s$, which we record here with the normalising constant that was immaterial there and is essential below,
\begin{equation}\label{eq:app_A:factor}
  \ell_{s,j}(a_{(j)}) = \frac{1}{\sqrt{2\pi\lambda_{s,j}}} \exp\!\left( -\frac{r_{s,j}^2}{2\lambda_{s,j}} \right), \qquad
  r_{s,j} = v_{s,j} + a_{(j)}^\top v_{s,1:j-1},
\end{equation}
with residuals $v_{s,l}=y_{s,l}-\Pi_{(l)}^\top z_s$. Since $r_{s,j}=A_{(j)}^\top v_s$, this is the same residual that the CSMC step of Section~\ref{sec:relevance:our:model} takes as its observation. It is convenient to read $\ell_{s,j}$ as an ordinary Gaussian regression in which $v_{s,j}$ is the \emph{response} and $v_{s,1:j-1}$ the \emph{regressors}. A missing $y_{t,l}$ is then a missing $v_{t,l}$.

\paragraph{Two cases.}
For a row $j\leq k$ the response $v_{t,j}$ and all regressors $v_{t,1:j-1}$ carry indices $\leq k$ and are observed, so $\ell_{t,j}$ can be evaluated and enters the full conditional exactly as in the complete-data case. For a row $j>k$ the period-$t$ factor is unusable as written: the response $v_{t,j}$ is missing, and for $j>k+1$ the regressor vector $v_{t,1:j-1}$ further mixes the observed $v_{t,1:k}$ with the missing $v_{t,k+1:j-1}$ inside a single square. The correct object is then the \emph{observed-data} likelihood, that is, the complete-data likelihood with the missing block marginalised out,
\begin{equation}\label{eq:app_A:obs}
  p(y_{t,1:k}\mid\theta,\lambda_t,z_t) 
  = \underbrace{\prod_{j=1}^{k}\ell_{t,j}(a_{(j)})}_{\text{only }a_{(1)},\dots,a_{(k)}}
  \;\times\;
  \underbrace{\int_{\mathbb{R}^{K-k}}\prod_{j=k+1}^{K}\ell_{t,j}(a_{(j)})\; \mathrm{d}y_{t,k+1:K}}_{\text{the coupled missing block}},
\end{equation}
the first $k$ factors pulling out because they do not involve the missing block. The identity covers both endpoints, the first product being empty at $k=0$ and the integral being empty at $k=K$.

\paragraph{The missing block integrates to $1$.}
For each missing row $j>k$ there are in fact two unobserved period-$t$ quantities: the observation $y_{t,j}$ and the latent log-volatility $\lambda_{t,j}$, the latter informed at the current date only by its own transition \eqref{eq:same_motion_model}, $\log\lambda_{t,j}=\log\lambda_{t-1,j}+e_{t,j}$, with $e_{t,j}\sim\mathcal{N}(0,\Phi_{jj})$. A correct observed-data likelihood marginalises both, which we do in two stages.

\emph{Stage 1 (fix the volatilities).} Hold arbitrary values $\lambda_{t,k+1:K}$ and change variables from the missing observations $y_{t,k+1:K}$ to the structural shocks $\epsilon_{t,j}=\lambda_{t,j}^{-1/2}r_{t,j}$, $j>k$. Two structural facts make this clean. Because $A$ is lower-triangular, a missing $y_{t,j}$ with $j>k$ enters only its own equation and later missing equations, never an observed row $j'\leq k$,  whose response and regressors all carry indices at most $j'\le k$. The missing observations are thus confined to the block $\{j>k\}$. Because $A$ has unit diagonal, $\partial\epsilon_{t,j}/\partial y_{t,j}=\lambda_{t,j}^{-1/2}$ and the Jacobian of $(\epsilon_{t,k+1:K})$ with respect to $(y_{t,k+1:K})$ is lower triangular with determinant $\prod_{j>k}\lambda_{t,j}^{-1/2}$. The substitution therefore contributes $\mathrm{d}y_{t,k+1:K}=\prod_{j>k}\lambda_{t,j}^{1/2}\,\mathrm{d}\epsilon_{t,k+1:K}$, which cancels the product of normalising constants in \eqref{eq:app_A:factor}, and
\begin{equation}\label{eq:app_A:J}
  J(\lambda_{t,k+1:K})
  \coloneq \int_{\mathbb{R}^{K-k}}\prod_{j=k+1}^{K}\ell_{t,j}(a_{(j)})\;
    \mathrm{d}y_{t,k+1:K}
  = \int \prod_{j=k+1}^{K}\frac{1}{\sqrt{2\pi}}\,e^{-\epsilon_{t,j}^2/2}\;
    \mathrm{d}\epsilon_{t,k+1:K}
  = 1
\end{equation}
for \emph{every} value of the volatilities. Each $\lambda_{t,j}$ enters only as the scale of its own shock, and that scale is undone by the very Jacobian factor needed to keep the term a proper density.

\emph{Stage 2 (integrate the volatilities out).} Since $J\equiv 1$ does not depend on $\lambda_{t,k+1:K}$, averaging it against any distribution $q(\lambda_{t,k+1:K})$, be it the prior transition, a point mass at a sampled value, or anything else, returns the same constant, $\int J(\lambda_{t,k+1:K})\,q(\lambda_{t,k+1:K})\,\mathrm{d}\lambda_{t,k+1:K}=1$. The unobserved volatilities of the missing rows therefore cannot influence the
$A$-update, however the sampler chooses to treat them.

Consequently the observed block is Gaussian, $p(y_{t,1:k}\mid\theta) = \mathcal{N}\!\big((\Pi z_t)_{1:k},\,\Omega_{1:k,1:k}\big)$, and because $A^{-1}$ is lower-triangular its top-left block inverts on its own,
\begin{equation*}
  \Omega_{1:k,1:k} = (A_{1:k,1:k})^{-1}\,\Lambda_{t,1:k}\,(A_{1:k,1:k})^{-\top},
\end{equation*}
involving only the leading block $A_{1:k,1:k}$ and the leading volatilities $\lambda_{t,1:k}$. There is no channel through which $\lambda_{t,k+1:K}$ or
$a_{(k+1)},\dots,a_{(K)}$ could enter the observed data.

\paragraph{Truncation is exact.}
By \eqref{eq:app_A:obs} and \eqref{eq:app_A:J}, and by Stage 2, period $t$ contributes a constant, free of $a_{(k+1)},\dots,a_{(K)}$, to the $A$-likelihood. The exact full conditional of $a_{(j)}$ for $j>k$ is thus the complete-data one of Appendix~\ref{app:updateA} with the data sum stopped one period early,
\begin{equation*}
  \bar{\Sigma}_{A,(j)}^{-1} = \Sigma_{A,(j)}^{-1} + \sum_{s=1}^{t-1}\frac{v_{s,1:j-1}\,v_{s,1:j-1}^\top}{\lambda_{s,j}}, \qquad
  \bar{\mu}_{A,(j)} = \bar{\Sigma}_{A,(j)}\!\left(\Sigma_{A,(j)}^{-1}\mu_{A,(j)}
    - \sum_{s=1}^{t-1}\frac{v_{s,j}\,v_{s,1:j-1}}{\lambda_{s,j}}\right), \quad j>k,
\end{equation*}
the sum being empty at $t=1$, in which case the full conditional is the prior. The observed rows $j\le k$ keep the full sum up to $t$, since their period-$t$ factors are evaluable and enter as in the complete-data case. This is exactly the per-row truncation already used for the $\wtl{\Pi}$ block. The Gaussian prior remains conjugate to the truncated likelihood, and no completion of $y_t$ is required.

\subsection{Update of \texorpdfstring{$\Phi$}{Phi} (Step 3)}\label{app:updatePhi}
Since $\Phi = \diag(\Phi_{1,1},\dots,\Phi_{K,K})$ is diagonal, its entries are
updated independently. From the transition~\eqref{eq:same_motion_model}, the
log-volatility increments $\Delta_{t,j} \coloneq \log\lambda_{t,j} -
\log\lambda_{t-1,j}$ are, conditionally on $\Phi_{j,j}$, independent
$\mathcal{N}(0,\Phi_{j,j})$ for $t=1,\dots,T$. Combined with the conjugate
inverse-Gamma prior $\Phi_{j,j}\sim\mathcal{I}\Gamma\bigl((K+2)/2,\,1/2\bigr)$ of
Section~\ref{sec:model}, the full conditional is
\begin{equation*}
  \pi(\Phi_{j,j}\mid \lambda_{1:T,j})
  \propto \Phi_{j,j}^{-(K+2)/2 - 1}\exp\left\{-\frac{1}{2\Phi_{j,j}}\right\}
  \prod_{t=1}^T \Phi_{j,j}^{-1/2}
  \exp\left\{-\frac{\Delta_{t,j}^2}{2\Phi_{j,j}}\right\},
\end{equation*}
which is again inverse-Gamma:
\begin{equation*}
  \Phi_{j,j}\mid \lambda_{1:T,j}
  \sim \mathcal{I}\Gamma\left( \frac{K+2+T}{2},\
  \frac{1 + \sum_{t=1}^T \Delta_{t,j}^2}{2} \right).
\end{equation*}
Here the first increment $\Delta_{1,j} = \log\lambda_{1,j} - \log\lambda_{0,j}$
involves the latent initial state $\lambda_{0,j}$ of~\eqref{eq:lambda0_prior},
so the sum runs over all $T$ increments.

\renewcommand{\refname}{References for the Appendix}
\begin{singlespace}
\putbib
\end{singlespace}
\end{bibunit}
\end{document}